\documentclass[journal,10pt]{IEEEtran}
\usepackage{algorithm,algorithmic}
\usepackage{amsfonts,enumitem}
\usepackage{algorithmic}
\usepackage{textcomp}
\def\BibTeX{{\rm B\kern-.05em{\sc i\kern-.025em b}\kern-.08em
		T\kern-.1667em\lower.7ex\hbox{E}\kern-.125emX}}

\usepackage[dvips]{graphicx}
\usepackage{epsfig,color}
\usepackage{amsmath,amssymb,multirow}
\usepackage{setspace}
\usepackage{listings}
\usepackage{cite}
\usepackage{stackengine}

\newcommand{\pp}{{\mathbf p}}

\newcommand{\tww}{{\mathbf {\tilde{w}}}}
\newcommand{\xx}{\mathbf{x}}

\newcommand{\zz}{{\mathbf z}}
\newcommand{\zj}{\mathbf {z}^J}
\newcommand{\tqq}{{\mathbf {\tilde{q}}}}
\newcommand{\tq}{{\tilde{q}}}
\newcommand{\tJ}{\mathbf{\tilde{J}}}
\newcommand{\yy}{\mathbf{y}}
\newcommand{\bphi}{\boldsymbol{\varphi}}

\newcommand{\bq}{{\boldsymbol{q}}}
\newcommand{\bv}{{\boldsymbol{v}}}

\newcommand{\tr}{\mathrm{tr}}

\newcommand{\bz}{\mathbf{z}^E}

\newcommand{\bw}{\mathbf{w}}
\newcommand{\bx}{\mathbf{x}}
\newcommand{\by}{\mathbf{y}}
\newcommand{\bs}{\mathbf{s}}
\newcommand{\btheta}{\boldsymbol{\theta}}
\newcommand{\bkappa}{\boldsymbol{\kappa}}
\newcommand{\bX}{\mathbf{X}}
\newcommand{\bY}{\mathbf{Y}}
\newcommand{\bl}{\boldsymbol{\lambda}}
\newcommand{\bL}{\boldsymbol{\Lambda}}
\newcommand{\bLd}{\boldsymbol{\Delta\Lambda}}
\newcommand{\bld}{\boldsymbol{\Delta\lambda}}
\newcommand{\blh}{\hat{\boldsymbol{\lambda}}}
\newcommand{\bLh}{\hat{\boldsymbol{\Lambda}}}

\usepackage{mathtools}

\newcommand\abs[1]{\left|#1\right|}
\newcommand\norm[1]{\left\lVert #1\right\rVert}

\begin{document}

	\title{Eavesdropper and Jammer Selection in Wireless Source Localization Networks}
	
	\author{Cuneyd Ozturk, \emph{Student Member, IEEE}, and Sinan Gezici,\thanks{The authors are with the Department of Electrical and Electronics Engineering, Bilkent University, Ankara, 06800, Turkey, E-mails:	\{cuneyd,gezici\}@ee.bilkent.edu.tr} \emph{Senior Member, IEEE}\thanks{Part of this work {was} presented at IEEE International Conference on Communications (ICC), June 2020 \cite{CuneydICC}.}\vspace{-0.5cm}}

	
	\maketitle
	
	\begin{abstract}
    We consider a wireless source localization network in which a target node emits localization signals that are used by anchor nodes to estimate the target node position. In addition to target and anchor nodes, there can also exist eavesdropper nodes and jammer nodes which aim to estimate the position of the target node and to degrade the accuracy of localization, respectively. We first propose the problem of eavesdropper selection with the goal of optimally placing a given number of eavesdropper nodes to a subset of possible positions in the network to estimate the target node position as accurately as possible. As the performance metric, the Cram\'{e}r-Rao lower bound (CRLB) related to the estimation of the target node position by eavesdropper nodes is derived, and its convexity and monotonicity properties are investigated. By relaxing the integer constraints, the eavesdropper selection problem is approximated by a convex optimization problem and algorithms are proposed for eavesdropper selection. Moreover, in the presence of parameter uncertainty, a robust version of the eavesdropper selection problem is developed. Then, the problem of jammer selection is proposed where the aim is to optimally place a given number of jammer nodes to a subset of possible positions for degrading the localization accuracy of the network as much as possible. A CRLB expression from the literature is used as the performance metric, and its concavity and monotonicity properties are derived. Also, a convex optimization problem and its robust version are derived after relaxation. Moreover, the joint eavesdropper and jammer selection problem is proposed with the goal of placing certain numbers of eavesdropper and jammer nodes to a subset of possible positions. Simulation results are presented to illustrate performance of the proposed algorithms.
	\end{abstract}
	
	
\vspace{-0.3cm}
	
\section{Introduction}\label{sec:Intro}

\subsection{Literature Review}

In wireless localization networks, position information is commonly extracted based on signal exchanges between anchor nodes with known positions and target (source) nodes whose position are to be estimated \cite{Zekavat,MoeWinNetworkOperation}. Based on the signaling procedure, wireless localization networks are classified into two groups as \textit{self localization} and \textit{source (network-centric) localization} networks \cite{Zekavat}. In the self localization scenario, target nodes estimate their positions via signals transmitted from anchor nodes whereas in source localization networks, anchor nodes estimate positions of target nodes from signals emitted by target nodes.
	
Wireless localization networks can be vulnerable to various attacks such as eavesdropping, jamming, sybil, and wormhole attacks \cite{SecureLoc,TheoreticalFoundation,LocPriv,TCOM2016}. For example, eavesdropper nodes may listen to signals transmitted from target nodes and estimate their positions, which breaches location secrecy \cite{TheoreticalFoundation,LocPriv}. In wireless localization networks, location secrecy cannot be guaranteed via encryption since location related information can be gathered by eavesdropper nodes by  just listening to signal exchanges rather than intercepting packets \cite{LocPriv}. As another type of attack, jammer nodes can degrade the localization accuracy of a network by transmitting jamming signals \cite{TCOM2016}. If jamming levels exceed certain limits, location information can be useless for specific applications due to its inaccuracy. In this manuscript, the focus is on eavesdropping and jamming attacks in wireless source localization networks.
	
In the literature, there exist only a few studies related to physical-layer location secrecy or eavesdropping in wireless localization networks \cite{TheoreticalFoundation,LocPriv,LocSec}. In \cite{TheoreticalFoundation}, a location secrecy metric (LSM) is proposed by considering only the position of a target node and the measurement model of an eavesdropper node. The aim of the eavesdropper node is to obtain an estimate of the target node position based on its measurement model, where the estimate can be either a point or a set of points. The definition of the LSM is based on the escaping probability of the target node from the eavesdropper node, i.e., the probability that the position of the target node is not an element of the set of estimated positions by the eavesdropper node. 
In practice, the measurement model of an eavesdropper node depends on several parameters in addition to the position of the target node \cite{LocSec}. For example, an eavesdropper node can extract location information based on signal exchanges between target and anchor nodes by using time difference of arrival (TDOA) approaches. In that case, the time offset becomes another unknown parameter. Hence, the definition of the LSM is extended in \cite{LocSec} by also taking channel conditions and time offsets into account. For some specific scenarios, LSM is calculated and algorithms are proposed to protect location secrecy by diminishing the estimation capability of an eavesdropper node \cite{LocSec}.
In \cite{LocPriv}, considering round-trip-measurements in a network, an eavesdropping model is presented by using TDOA approaches. Also, power allocation frameworks for anchor and target nodes are presented to degrade the estimation performance of an eavesdropper node while maintaining the localization accuracy of the network \cite{LocPriv}.
	
Related to jamming and anti-jamming techniques in wireless localization networks, a great amount of research has been conducted in the literature
\cite{GroverJam14,Vadlamani2014ABP,GeziciJamPlacement16,JammingWCL,HoujeijJam14,LiJam10,LiuJam12,SunJam09,ChengJam11,HusseinSurveyJam15,FengJamPlacement14,TCOM2016,SankararamanJam12,SezerPowerControl18, BayramJam17}. Placement of jammer nodes in wireless localization networks can serve for different purposes \cite{Vadlamani2014ABP}. Namely, the aim of placing jammer nodes can be either to reduce the localization accuracy of the network (i.e., adversarial) \cite{GeziciJamPlacement16,JammingWCL,TCOM2016, ShankarJam08,BayramJam17}, or to protect the network from eavesdropper attacks \cite{SankararamanJam12,GroverJam14, HusseinSurveyJam15, ChengJam11, SunJam09, LiuJam12, LiJam10, HoujeijJam14}. In \cite{TCOM2016}, optimal power allocation schemes are developed for jammer nodes under peak and total power limits by maximizing the average or minimum Cram\'{e}r-Rao lower bounds (CRLBs) in self localization networks. The same problem is considered in \cite{JammingWCL} for source localization networks. In \cite{BayramJam17}, the average CRLB of target nodes is maximized while keeping their minimum CRLB above a certain threshold for self-localization networks. In \cite{TCOM2016,JammingWCL,BayramJam17}, it is assumed that positions and the number of jammer nodes are fixed. When positions of jammer nodes can be changed, their optimal placement can be considered for achieving the best jamming performance. In \cite{GeziciJamPlacement16}, the optimal jammer placement problem is investigated for wireless self-localization networks in the presence of constraints on possible locations of jammer nodes. On the other hand, in \cite{SankararamanJam12}, jammer nodes are placed to reduce the received signal quality of eavesdropper nodes while not preventing the operation of the actual network.
	
Game theoretic approaches are also utilized for determining jamming strategies	\cite{Vadlamani2014ABP, SezerPowerControl18}. In \cite{Vadlamani2014ABP}, an attacker tries to maximize the damage on network activity while the aim of a defender is to secure a multi-hop multi-channel network. The action of the attacker is determined by the selection of jammer node positions and a channel hopping strategy whereas the action of the defender is based on the channel hopping strategy. In \cite{SezerPowerControl18}, two different power control games between anchor nodes and jammer nodes are formulated for self-localization networks based on the average CRLB and the worst-case CRLB criteria. Nash equilibria of the proposed games are analyzed and it is shown that both games have at least one pure-strategy Nash equilibrium.
	
In the literature, eavesdropping and jamming attacks have not been considered jointly for wireless localization networks. However, for communications networks, \cite{LiuActiveEav19,MukherjeeEavJam13,ZhuEavJam11} investigate effects of jamming and eavesdropping together. In \cite{LiuActiveEav19}, a secure transmission scheme is proposed for a wiretap channel when a source communicates with a legitimate unmanned aerial vehicle (UAV) in the presence of eavesdroppers. Full duplex active eavesdropping is assumed, i.e., wiretappers can perform eavesdropping and jamming simultaneously. In \cite{MukherjeeEavJam13}, a multiple-input multiple-output communication system with a transmitter, a receiver and an adversarial wiretapper is considered. The wiretapper is able to act as either an eavesdropper or a jammer. The transmitter makes a decision between allocating all the power to information signals or broadcasting some artificial interference signals to jam the wiretapper. A game theoretic formulation of this problem is also given in \cite{MukherjeeEavJam13}, and its Nash equilibria are analyzed. {In \cite{ZhuEavJam11}, the considered wireless network contains wireless users, relay stations, base station (BS), and an attacker who has the ability to act as an eavesdropper and as a jammer. The aim of the attacker is to degrade the secrecy rate of the network and the transmission rate of the users. Each user connects to one of the relay stations so that the amount of potential interference from other users is reduced and the expected level of security for the transmission is increased. This problem is formulated as an $(N+1)$ person noncooperative game where $N$ is the number of users and existence of mixed-strategy Nash equilibria is shown.}
	
\vspace{-0.2cm}

\subsection{Contributions}

Although a location secrecy metric is developed in \cite{TheoreticalFoundation,LocSec} and the problem of protecting location secrecy is investigated in \cite{LocPriv}, there exist no studies that consider the problem of \textit{eavesdropper selection}. In the proposed eavesdropper selection problem, the aim is to optimally place a given number of eavesdropper nodes to a subset of possible positions such that the location secrecy of target nodes is reduced as much as possible. The optimal eavesdropper selection problem is studied from the perspective of eavesdropper nodes for determining performance limits of eavesdropping. {The CRLB for estimation of target node positions by eavesdroppers is employed as the performance metric.}
The eavesdropper selection problem also carries similarities to the anchor placement problem (e.g., \cite{IsaacsSensor10,XuSensor19,LiuNodePlacement17}), in which the aim is to determine the optimal positions of anchor (reference) nodes for optimizing accuracy of target localization. While the optimization is performed over positions of anchor nodes in the anchor placement problem, the aim is to choose the best positions from a finite set of possible positions in the eavesdropper selection problem. (Hence, different theoretical approaches are utilized in this manuscript.)

In addition, even though jamming and anti-jamming strategies are investigated extensively under various scenarios in \cite{GroverJam14,Vadlamani2014ABP,GeziciJamPlacement16,JammingWCL,HoujeijJam14,LiJam10,LiuJam12,SunJam09,ChengJam11,HusseinSurveyJam15,FengJamPlacement14,TCOM2016,SankararamanJam12,SezerPowerControl18, BayramJam17}, there has been no consideration about \textit{jammer selection}. In the proposed jammer selection problem, the goal is to place a given number of jammer nodes to a subset of possible positions to degrade the localization accuracy of a wireless network {where the CRLB related to estimation of target node positions by anchor nodes is used as the performance metric.}

Moreover, despite the work in \cite{LiuActiveEav19,MukherjeeEavJam13,ZhuEavJam11}, which consider both jamming and eavesdropping for wireless communication networks {based on performance metrics such as outage probability, transmission rate and secrecy rate}, the presence of jammer and eavesdropper nodes together has not been investigated for wireless localization networks. In this manuscript, we focus on a wireless localization network with multiple eavesdropper and jammer nodes, and formulate the \textit{joint eavesdropper and jammer selection} problem {by employing the CRLB as an estimation theoretic performance metric.} The goal is to place certain numbers of eavesdropper and jammer nodes to a subset of possible positions in order to degrade the accuracy of the localization network while keeping the eavesdropping capability above a threshold. {In particular, eavesdropper nodes aim to minimize the average CRLB related to their estimation of target node positions whereas jammer nodes seek to maximize the average CRLB for estimating target node positions by anchor nodes via emitting noise signals.}

The main contributions of this manuscript can be specified as follows:
\begin{itemize}[leftmargin=*]
\item We formulate the eavesdropper selection, jammer selection, and joint eavesdropper and jammer selection problems in a wireless source localization network for the first time in the literature.
\item For the eavesdropper selection problem, a novel CRLB expression (used as a performance metric for location secrecy) is derived related to the estimation of target node positions by eavesdropper nodes (Proposition~1).
\item We prove that the CRLB expression derived for the eavesdropper selection problem is convex and non-increasing with respect to the selection vector, which specifies the selection of positions for placing eavesdropper nodes (Proposition~2 and Lemma 1).
\item For the jammer selection problem, we utilize a CRLB expression from the literature and prove that it is concave and non-decreasing with respect to the selection vector (Proposition~3 and Lemma 3).
\item We express the eavesdropper selection, jammer selection, and joint eavesdropper and jammer selection problems as convex optimization problems after relaxation.
\item We propose algorithms to solve the proposed problems by considering both perfect and imperfect knowledge of system parameters, and develop robust approaches in the presence of imperfect knowledge.
    \end{itemize}

In the conference version of this manuscript \cite{CuneydICC}, only the
eavesdropper selection problem is considered with shorter proofs of
propositions and without proofs of lemmas. In this manuscript,
{the eavesdropper selection problem is investigated by
	providing complete proofs for all theoretical results and performing extensive simulations over a large network. In addition, the jammer selection and the joint eavesdropper
	and jammer selection problems are proposed and analyzed. Although the
	CRLB expression for the jammer selection problem is taken from the
	literature, its concavity and monotonicity properties are derived for
	the first time in the literature. Based on these properties, convexity
	of a jammer power allocation problem in the literature is also implied
	and a robust jammer selection problem is formulated.}


\vspace{-0.2cm}

\subsection{{Motivation}}\label{sec:Moti}

The investigation of the eavesdropper selection, jammer selection, and joint eavesdropper and jammer selection problems is important to identify the adversarial capabilities of eavesdropper and/or jammer nodes.

{
As a motivating example of an application scenario for the eavesdropper selection problem, consider
a restricted environment such as a military facility or a factory (e.g.,
imagine an area in Fig.~\ref{fig:Network} covering blue squares and cross signs). In this environment, target nodes can represent the personnel or important
equipment, which send signals to anchors nodes so that their locations
can be tracked by the wireless localization network. A fixed number of
eavesdropper nodes can be placed at some of feasible locations outside
the restricted environment (red triangles in Fig.~\ref{fig:Network}), e.g., under some camouflage. The aim of eavesdropper nodes is to gather accurate location
information about target nodes (i.e., personnel or equipment) for
leaking critical information. To this aim, they need to be placed at
optimal locations among the feasible locations, leading to the proposed
eavesdropper selection problem.}

{
Considering the same setting, jammer nodes can be placed at some of
feasible locations for the purpose of reducing the accuracy of the
localization network so that the network will not be able to track
critical equipment or personnel with sufficient localization accuracy.
This scenario can also be encountered in a battle-field in order to
disrupt the localization capability of an enemy network. Similarly, the
joint eavesdropper and jammer selection problem can be considered for
both gathering location information about target nodes and reducing the
accuracy of the localization network.}

\vspace{-0.2cm}

\subsection{Notation}

    Throughout the manuscript, $\bX \succeq \bY$ denotes that  $\bX-\bY$ is a positive semi-definite matrix, $\bx \succeq \by $ means that $x_i \geq y_i$ for all $i = 1,2, \ldots, n$, where $\bx = [x_1 ~ x_2 \ldots x_n]^\intercal$ and  $\by  = [y_1 ~ y_2 \ldots  y_n]^\intercal$, and $\tr\{\cdot\}$ represents the trace of a square matrix. Also, the following definitions are used: $(i)$ Let $f(\cdot)$ be a real-valued function of $\zz\in\mathbb{R}^{n}$. $f(\zz)$ being non-increasing in $\zz$ means that if $\zz$ and $\bw$ satisfy $\zz \succeq \bw$, $f(\zz)\leq f(\bw)$ holds.
    $(ii)$ Let $g(\cdot)$ be a real-valued of function of $\bX\in S^n_+$, where $S^n_+$ is the set of positive semi-definite matrices in $\mathbb{R}^{n\times n}$. Then, $g(\bX)$ being non-increasing in $\bX$ means that if $\bX$ and $\bY$  satisfy $\bX \succeq \bY$, $g(\bX)\leq g(\bY)$ holds.

\vspace{-0.2cm}
	
	\section{System Model}\label{sec:SysModel}
	
	Consider a two-dimensional wireless source localization network in which a target node (source) transmits signals that are used by anchor nodes to estimate its location. The number of anchor nodes is denoted by $N_A$ and they are located at $\yy_{j}\in\mathbb{R}^{2}$ for $j = 1, 2, \ldots, N_A$. Also, there exists some prior information about the location of the target node such that it is located at $\xx_i\in\mathbb{R}^2$ with probability $w_i\geq 0$ for $i = 1, 2, \ldots, N_T$, where $N_T$ is the number of possible locations for the target node, and $\sum_{i=1}^{N_T} w_i = 1$. Let $\mathcal{A}_i$ represent the set of locations of anchor nodes that are connected to the $i$th target position (i.e., location $\xx_i$) for $i = 1, 2, \ldots, N_T$. Moreover, let $\mathcal{A}_L^{(i)}$ and $\mathcal{A}_{NL}^{(i)}$ denote, respectively, the locations of anchor nodes having line-of-sight (LOS) and non-line-of-sight (NLOS) connections to the target node located at $\xx_i$.
	
	In the wireless localization network, there also exist $N$ different locations specified by the set $\mathcal{N} = \{\pp_{1}, \pp_{2}, \ldots, \pp_{N}\}$, at which either jammer or eavesdropper nodes can be placed. Eavesdropper nodes listen to the signals transmitted from the target node to the anchor nodes and aim to estimate the location of the target node. On the other hand, jammer nodes degrade the localization performance of the anchor nodes by transmitting zero-mean white Gaussian noise \cite{TCOM2016,jammernoise}. It is assumed that at any given time, at most $N_E$ locations in $\mathcal{N}$ can be used for eavesdropping purposes, whereas at most $N_J$ of them can be used for jamming purposes, where $N_E+N_J \leq N$.  In other words, there exist at most $N_E$ eavesdropper nodes and $N_J$ jammer nodes that can be placed at some of the $N$ possible locations. Let $\mathcal{N}_{E}$ and $\mathcal{N}_{J}$ denote the set of locations in $\mathcal{N}$ at which eavesdropper nodes and jammer nodes are placed, respectively.

{Considering a wideband wireless localization network as in \cite{part1},} 	
the signal transmitted from the $i$th target position (i.e., $\xx_i$) that is intended for the anchor node located at $\yy_j$ is denoted by $s_{ij}(t)$. If an eavesdropper node is placed at $\pp_{k}$ (i.e., if $\pp_k\in\mathcal{N}_E$), the received signal at that eavesdropper node due to the
transmission of $s_{ij}(t)$
is represented by $r_{ijk}^{E}(t)$. This signal is expressed as
\begin{equation}\label{eq:recwaveform}
	r_{ijk}^{E}(t) = \sum_{l=1}^{L_{ijk}^{E}}\alpha_{ijk}^{(E,l)} s_{ij}\big(t-\tau_{ijk}^{(E,l)}\big) +n_{ijk}(t)
	\end{equation}
for $t\in[T_{1}^{(E,k)},T_{2}^{(E,k)})$ and $(i,j)\in\mathcal{S}_{k}$, where $T_{1}^{(E,k)}$ and $T_{2}^{(E,k)}$ specify the observation interval for the eavesdropper node located at $\pp_k$, $\mathcal{S}_k = \{(i,j)\mid \pp_k\in\mathcal{N}_E,\, \yy_j\in\mathcal{A}_i\}$, $L_{ijk}^{E}$ represents the number of paths between the target node located at $\xx_i$ and the eavesdropper node located at $\pp_k$ (due to the
transmission of $s_{ij}(t)$),
$\alpha_{ijk}^{(E,l)}$ and $\tau_{ijk}^{(E,l)}$ denote, respectively, the amplitude and the delay of the $l$th multipath component, and
$n_{ijk}(t)$ is zero-mean white Gaussian noise with a power spectral density level of $\sigma_{k}^2$. Considering orthogonal channels between target and anchor nodes, $n_{ijk}(t)$ is modeled as independent for all $i,j,k$ \cite{GeziciJamPlacement16,JammingWCL,PowerOpt}. The delays of the paths are characterized by the following expression:
	\begin{equation} \label{eq:delayeq}
	\tau_{ijk}^{(E,l)} = \frac{1}{c} \left(\norm{\xx_i-\pp_k} + b_{ijk}^{(E,l)} + \Delta_i \right)
	\end{equation}
where $c$ is the propagation speed, $b_{ijk}^{(E,l)}\geq 0$ is the range bias ($b_{ijk}^{(E,1)} = 0$ for LOS propagation and $b_{ijk}^{(E,1)}>0$ for NLOS), and $\Delta_i$ characterizes the time offset between the clocks of the target node located at $\xx_i$ and the eavesdropper nodes. {It is assumed that the eavesdropper nodes are perfectly synchronized among themselves and there exist no clock drifts. (Please see \cite{SyncMSThesis,SyncPhDThesis} for clock drift mitigation mechanisms.)} However, there is no synchronization between the target node and the eavesdropper nodes. Furthermore, for any $i = 1, 2, \ldots, N_T$, we define $\mathcal{N}^{(i)}_L \triangleq \{(j,k)\mid\ b_{ijk}^{(E,1)} = 0\}$ and $\mathcal{N}^{(i)}_{NL} \triangleq \{(j,k)\mid\ b_{ijk}^{(E,1)} \neq 0\}$, which are the set of anchor and eavesdropper node indices corresponding, respectively, to LOS and NLOS connections between the eavesdropper nodes and the target node located at $\xx_i$. (For example, if $b_{i32}^{(E,1)} = 0$, it means that the eavesdropper node at position $\pp_2$ and the target node at position $\xx_i$ are in LOS during the transmission of the signal from that target node to the anchor node at position $\yy_3$ (i.e., during the transmission of $s_{i3}(t)$).)
	
	On the other hand, due to the existence of jammer nodes, the signal received at the anchor node located at $\yy_j$ coming from the target node located at $\xx_i$ can be expressed as
	\begin{equation}\label{eq:recwaveform2}
	\hspace{-0.05cm}{r_{ij}^{A}(t) = \sum_{l=1}^{{L}_{ij}^{A}}\alpha_{ij}^{(A,l)} s_{ij}\big(t-\tau_{ij}^{(A,l)}\big)} + \hspace{-0.1cm} \sum_{\{l:\pp_{l}\in\mathcal{N}_{J}\}} \hspace{-0.25cm}\gamma_{lj} \sqrt{P_{l}^{J}} v_{lij}(t)+ \eta_{ij}(t)
	\end{equation}
for the observation interval ${[T_{1}^{(A,j)},T_{2}^{(A,j)})}$ and for $\yy_j\in\mathcal{A}_i$, where ${\alpha_{ij}^{(A,l)}}$ and ${\tau_{ij}^{(A,l)}}$ denote, respectively, the amplitude and the delay of the $l$th multipath component between the target node at location $\xx_i$ and the anchor node at location $\yy_j$, ${{L}_{ij}^{A}}$ represents the number of multipaths between the target node at location $\xx_i$ and anchor node at location $\yy_j$, $\gamma_{lj}$ is the channel coefficient between the anchor node at location $\yy_j$ and the jammer node located at $\pp_{l}$, and $P_{l}^{J}$ is the transmit power of the jammer node at position $\pp_{l}$. Moreover, $\sqrt{P_{l}^{J}} v_{lij}(t)$ and $ \eta_{ij}(t)$ are the jammer noise and the measurement noise, respectively. It is assumed that both of them are independent zero-mean white Gaussian random processes, where the  average power of $v_{lij}(t)$ is equal to one and that of $ \eta_{ij}(t)$ is equal to $\tilde{\sigma_j}^2$. It is modeled that $v_{lij}(t)$ is independent for all $l, i, j$ and $\eta_{ij}(t)$ is independent for all $i,j$ due to the presence of
orthogonal channels between target and anchor nodes \cite{JammingWCL}. Furthermore, the delays of the paths are characterized by
	\begin{equation} \label{eq:delayeq2}
	{\tau_{ij}^{(A,l)}} = \frac{1}{c} \left(\norm{\yy_j-\xx_i} + {b_{ij}^{(A,l)}}\right)
	\end{equation}
where ${b_{ij}^{(A,l)}}\geq 0$ is the range bias of the $l$th path between the target node located at $\xx_i$ and the anchor node located at $\yy_j$. ($ {b_{ij}^{(A,1)}} = 0$ for LOS propagation and ${b_{ij}^{(A,1)}}>0$ for NLOS.) {Unlike the expression in \eqref{eq:delayeq}, no clock offsets are considered in \eqref{eq:delayeq2} since target and anchor nodes are assumed to be synchronized.}
\vspace{-0.2cm}
\section{Eavesdropper Selection Problem} \label{sec:EavSel}
	
In this section, we assume that there exist only eavesdropper nodes in the environment, i.e., $N_J = 0$, and focus on the eavesdropper selection problem. In this case, the aim is to choose at most $N_E$ locations from set $\mathcal{N}$ for eavesdropping purposes so that the location of the target node is estimated as accurately as possible. 

For quantifying the location estimation accuracy, the CRLB is used as a performance metric since the mean-squared error of the maximum likelihood (ML) estimator is asymptotically tight to the CRLB in the high SNR regime \cite{DETEST_Poor}. Based on the CRLB metric, the eavesdropper selection problem is investigated in the presence of perfect and imperfect knowledge of system parameters in the following sections.

	\subsection{Problem Formulation}

	To formulate the eavesdropper selection problem, we introduce a selection vector $\bz = [z_1^{E} ~ z_2^{E} \ldots  z_{N}^{E}]^\intercal$, specified as
	\begin{equation}\label{eq:zj}
	z_k^{E}  = \begin{cases}
	1, &\text{if  $\pp_k\in\mathcal{N}_{E}$} \\
	0, &\text{otherwise}
	\end{cases}
	\end{equation}
	where $\sum_{k = 1}^{N} z_k^{E}  \leq N_E$. In addition, for the target position $i$, $\btheta_{i}$ is defined as follows:
	\begin{equation}
	\btheta_{i} \triangleq [\xx_i^\intercal ~ \Delta_i ~ \bkappa_{i1}^\intercal ~ \bkappa_{i2}^\intercal  \ldots  \bkappa_{iN}^\intercal ]^\intercal
	\end{equation}
	where $\bkappa_{ik}$ is the vector obtained by concatenating the elements of $\mathbf{\tilde{\bkappa}}_{ijk}$ vertically, $\bkappa_{ik} = [\tilde{\bkappa}_{ijk}^{\intercal}]^{\intercal}_{j\in\mathcal{A}_i}$, with
	\begin{equation*}
	\mathbf{\tilde{\bkappa}}_{ijk} = \begin{cases}
	[\alpha_{ijk}^{(E,1)} ~ b_{ijk}^{(E,2)} ~ \ldots  b_{ijk}^{(E,L_{ijk}^E)} ~ \alpha_{ijk}^{(E,L_{ijk}^E)} ]^\intercal, \text{if $b_{ijk}^{(E,1)} = 0$}
	\\
	[b_{ijk}^{(E,2)} ~ \alpha_{ijk}^{(E,2)}  \ldots  b_{ijk}^{(E,L_{ijk}^E)} ~ \alpha_{ijk}^{(E,L_{ijk}^E)} ]^\intercal, \text{otherwise.}
	\end{cases}
	\end{equation*}
	for any $i,j,k.$
	
	It is known that the estimation error vector satisfies \cite{DETEST_Poor}
	\begin{equation} \label{eq:crlb1}
	\mathbb{E}_{\btheta_{i}} \{(\btheta_{i}-\hat{\btheta}_i)(\btheta_{i}-\hat{\btheta}_i)^\intercal\}\succeq \mathbf{J}_{\btheta_i}^{-1}
	\end{equation}
	where $\hat{\btheta}_i$ is any unbiased estimate of $\btheta_i$, and $\mathbf{J}_{\btheta_i}$ is the Fisher information matrix (FIM) for the parameter vector $\btheta_i$. From \eqref{eq:crlb1}, the CRLB for estimating the position of the target node located at $\xx_i$ is obtained as
	\begin{equation}\label{eq:CRLBgeneric}
	\mathbb{E}_{\btheta_i} \{ \norm{\hat{\xx}_i-\xx_i}^2\} \geq \tr \{[\mathbf{J}_{\btheta_i}^{-1}]_{2\times 2}\}
	\end{equation}
	where $\hat{\xx}_i$ is any unbiased estimate of $\xx_i$.
	It is noted from \eqref{eq:CRLBgeneric} that, for the CRLB calculation, we should focus on the equivalent Fisher information matrix (EFIM) for $\xx_i$, which is a $2\times 2$ matrix denoted by $\mathbf{J}_e^{(i)}(\xx_i)$ such that $[\mathbf{J}_{\btheta_i}^{-1}]_{2\times 2} = \big(\mathbf{J}_e^{(i)}(\xx_i)\big)^{-1}$ \cite{part1}. Since $[\mathbf{J}_{\btheta_i}]_{2\times 2}$ is a function of both $\xx_i$ and $\bz$, it is convenient to write  $[\mathbf{J}_{\btheta_i}]_{2\times 2} \triangleq\mathbf{J}_{e}^{(i)}(\xx_i,\bz) $. Hence, we formulate the proposed eavesdropper selection problem as follows:
	\begin{subequations}\label{eq:opt}
		\begin{align}
		&\!\min_{\bz}        &\quad& \sum_{i=1}^{N_T} w_i \, \tr\big\{\big(\mathbf{J}_{e}^{(i)}(\xx_i,\bz)\big)^{-1} \big\} \label{eq:optProb}\\
		&\text{subject to} &      & \sum_{k=1}^{N} z_k^E \leq N_E,\label{eq:constraint1}\\
		&                  &      & z_k^E \in \{0,1\} \text{ for }  k = 1,2, \ldots, N  .\label{eq:constraint2}
		\end{align}
	\end{subequations}
Namely, the aim is to select the best locations for eavesdropper nodes for achieving the minimum average CRLB by considering possible target node positions ($\xx_i$) and their probabilities ($w_i$).

\subsection{Theoretical Results and Algorithms}

To simplify the notation, let $f(\bz)$ represent the objective function in \eqref{eq:opt}; that is,
	\begin{equation}\label{eq:fz}
	f(\bz) \triangleq \sum_{i=1}^{N_T} w_i \, \tr\big\{\big(\mathbf{J}_{e}^{(i)}(\xx_i,\bz)\big)^{-1} \big\}\,.
	\end{equation}
	In the rest of this section, we first obtain a closed form expression of $\tr\big\{\big(\mathbf{J}_{e}^{(i)}(\xx_i,\bz)\big)^{-1} \big\}$ for any target location $i$, and then analyze monotonicity and convexity properties of $f(\bz)$ with respect to $\bz$.
	
	\textit{Proposition 1:} For a given eavesdropper selection vector $\bz$, the CRLB for estimating the position of the target node located at $\xx_i$ is given by
	\begin{align}\label{eq:closedform}
	\tr\big\{\big(\mathbf{J}_{e}^{(i)}(\xx_i,\bz)\big)^{-1} \big\} = \frac{\tilde{p}_i(\bz)}{\tilde{r}_i{(\bz)}}
	\end{align}
	where
	\begin{align}\label{eq:ptil}
	&\tilde{p}_i(\bz)  = 3\sum_{(u,k)\in\mathcal{N}_L^{(i)}} \sum_{(v,l)\in\mathcal{N}_L^{(i)} } \hspace{-0.1cm} z_k^E z_l^E \lambda_{uk}^{(i)} \lambda_{vl}^{(i)} p_{k,l}^{(i)}, \\\label{eq:rtil}
	&\tilde{r}_i(\bz)  = 4\sum_{(u,k)\in\mathcal{N}_L^{(i)}}\sum_{(v,l)\in\mathcal{N}_L^{(i)}} \sum_{(s,m)\in\mathcal{N}_L^{(i)}} z_k^E z_l^E z_m^E \\
	&{\hspace{1cm}}\times\lambda_{uk}^{(i)} \lambda_{vl}^{(i)} \lambda_{sm}^{(i)} p_{k,l}^{(i)} p_{l,m}^{(i)} p_{m,k}^{(i)}, \\
	&\lambda_{jk}^{(i)}  = \frac{8\pi \beta_{ij}^2}{c^2}(1-\chi_{jk}^{(i)}) {\rm{SNR}}_{ijk}^{(1)}, \label{eq:lambdas}\\
	&\beta_{ij}^2  = \frac{\int_{-\infty}^{\infty} f^2 |S_{ij}(f)|^2\,df}{\int_{-\infty}^{\infty} |S_{ij}(f)|^2\,df}, \\
	&{\rm{SNR}}_{ijk}^{(1)} = \frac{|\alpha_{ijk}^{(E,1)}|^2 \int_{-\infty}^{\infty} |S_{ij}(f)|^2\,df}{2\sigma_k^2}, \\
	&p_{k,l}^{(i)} = \sin^2\bigg(\frac{\phi_{ik}-\phi_{il}}{2}\bigg)\label{eq:SNRs}
	\end{align}
with $S_{ij}(f)$ denoting the Fourier transform of $s_{ij}(t)$, $\chi_{jk}^{(i)}$ being the path overlap coefficient with $0\leq \chi_{jk}^{(i)} \leq 1$ \cite{part1}, and $\phi_{ik}$ representing the angle from the $i$th target location to $\pp_{k}$, i.e., $\phi_{ik} = \arctan{\frac{x_{i2}-p_{k2}}{x_{i1}-p_{k1}}}$ ($\xx_{i} = [x_{i1} ~x_{i2}]^\intercal$, $\pp_{k} = [p_{k1} ~p_{k2}]^\intercal$).
	
	\textit{Proof:} {See Appendix-A.\hfill $\blacksquare$}

{In Proposition~1, the CRLB is expressed in closed-form as a ratio of two polynomials in terms of the eavesdropper selection vector, which brings benefits in terms of computational cost. For example, it}	
facilitates the calculation of the solution of \eqref{eq:opt} via an exhaustive search over all possible $\bz$ vectors when $N$ is sufficiently small. Also, it is noted that the proposed CRLB expression in Proposition~1 depends only on the LOS signals (see \eqref{eq:closedform}--\eqref{eq:rtil}), which is in accordance with the results in the literature (e.g., \cite[Prop.~1]{part1} and \cite{YihongTez}).
	
\indent\textit{Remark 1:} It is observed from the CRLB expression in \eqref{eq:closedform}--\eqref{eq:rtil} that if all $\lambda_{jk}^{(i)}$'s are scaled by the same nonnegative real number $\xi$,  $\tr\big\{\big(\mathbf{J}_{e}^{(i)}(\xx_i,\bz)\big)^{-1} \big\}$ is scaled by $1/\xi$ for all $i = 1,2, \ldots, N_T$. Therefore, the optimal eavesdropper selection strategy (i.e., the solution of \eqref{eq:opt}) remains the same in such cases.

\indent{\textit{Remark 2:} For the eavesdropper selection problem, the \emph{probability distribution} of the target node positions is assumed to be known. Also, it is assumed that LOS/NLOS conditions for possible target-eavesdropper positions and $\lambda_{jk}^{(i)}$'s are known. Although these assumptions may not hold in some practical scenarios, they facilitate calculation of theoretical limits on the best achievable performance of eavesdropper nodes \cite{TCOM2016}. If eavesdropper nodes are smart and can learn all the environmental parameters, the localization accuracy derived in this work can be achieved; otherwise, the localization accuracy (hence the eavesdropping capability) is bounded by the obtained results.\footnote{{The tightness of the provided bounds in the presence of imperfect information about the distribution of the target node location is evaluated in Section~\ref{sec:SimNew}.}} {In addition, when the $\lambda_{jk}^{(i)}$ terms and LOS/NLOS conditions are not known perfectly, the robust formulation of the eavesdropper selection problem in Section \ref{sec:Robust} can be employed to provide a more practical formulation (please also see Remark~6).}}	


	The following lemma characterizes the monotonicity of $f(\bz)$ in \eqref{eq:fz} (i.e., the objective function in \eqref{eq:opt}) with respect to $\bz$, which is also utilized in the analysis in Section~\ref{sec:Robust} (Lemma~2).
	
	\textit{Lemma 1:} $f(\bz)$ is non-increasing in $\bz$.

	\textit{Proof:} {See Appendix-B.\hfill $\blacksquare$}
	
	This result is actually quite intuitive as one expects improved performance for estimating the location of a target node as the number of eavesdropper nodes increases.
	Next, we prove the convexity of the objective function in \eqref{eq:opt} with respect to $\bz$.
	
	\textit{Proposition 2:} $f(\bz)$ in \eqref{eq:fz} is a convex function of $\bz$.
	
	\textit{Proof:} {See Appendix-C.\hfill $\blacksquare$}
	
	As a consequence of Proposition~2, the optimization problem in \eqref{eq:opt} becomes a convex optimization problem by relaxing the last constraint in \eqref{eq:constraint2}. Furthermore, it is deduced from Lemma~1 that if $\mathbf{z}^{*}= [z^{*}_1 ~ z^{*}_2  \ldots z^{*}_N]^\intercal$ is a solution of \eqref{eq:opt}, then \eqref{eq:constraint1} must be satisfied with equality, i.e., $\sum_{j=1}^{N} z^{*}_j = N_{E}$ must hold. Therefore, the relaxed version of \eqref{eq:opt} can be formulated as follows:
	\begin{subequations}\label{eq:opt2}
		\begin{align}
		&\!\min_{\bz}        &\quad& \sum_{i=1}^{N_T} w_i \tr\big\{\big(\mathbf{J}_{e}^{(i)}(\xx_i,\bz)\big)^{-1} \big\} \label{eq:optProb2}\\
		&\text{subject to} &      & \sum_{k=1}^{N} z_k^E  = N_E,\label{eq:constraint21}\\
		&                  &      & 0 \leq z_k^E \leq 1 \text{ for }  k = 1,2, \ldots, N  .\label{eq:constraint22}
		\end{align}
	\end{subequations}
	
	As \eqref{eq:opt2} is a convex problem, its solution can be obtained via convex optimization tools \cite{boyd} {(called the \textit{relaxed algorithm} in Section~\ref{sec:Simu})}. After finding the solution of \eqref{eq:opt2}, we  propose the following two algorithms to obtain a solution of the original problem in \eqref{eq:opt}. First, we can simply set the largest $N_E$ components of the solution of \eqref{eq:opt2} to one, and the others to zero (called the \textit{largest-$N_E$ algorithm} in Section~\ref{sec:Simu}). Second, starting from this solution, we can use {a modified version of the Local Optimization} algorithm discussed in \cite{boydselection} and obtain the solution of \eqref{eq:opt} (called \textit{the proposed swap algorithm} in Section~\ref{sec:Simu}). {The details of the proposed swap algorithm is provided in Algorithm 1, where $\zz^{*}$ and $\zz^{*}_{\text{largest-}N_E}$ denote the optimal selection vectors obtained by the relaxed algorithm  and the largest-$N_E$ algorithm, respectively,} {$N_{\text{swap}}^{\text{max}}$ is the upper limit for the number of swap operations, and $\mu$ determines the stopping criterion.} {While performing one swap operation,} one checks whether there is a decrease in the objective function by simply swapping one of the $N_E$ selected positions with one of the $N-N_E$ positions that are not selected.
	
	\begin{algorithm}[ht]
	\caption{{Proposed Swap Algorithm}}
	{
		\begin{algorithmic}[1]
			\renewcommand{\algorithmicrequire}{\textbf{Input:}}
			\renewcommand{\algorithmicensure}{\textbf{Output:}}
			\REQUIRE $\zz^{*}, \zz^{*}_{\text{largest-}N_E}, \mu, N_{\text{swap}}^{\text{max}}$
			\ENSURE  $\zz^{*}_{\text{swap}}$.
			\STATE Set 	\textbf{boolean} b $\leftarrow$ true, c$\leftarrow$0
			\IF {$\lvert f(\zz^{*})-f(\zz^{*}_{\text{largest-}N_E}) \rvert \leq \mu f(\zz^{*})$}
			\STATE b $\leftarrow$ false, $\zz^{*}_{\text{swap}}\leftarrow \zz^{*}_{\text{largest-}N_E}$
			\ELSE
			\STATE $\zz_{\text{temp}}\leftarrow \zz^{*}_{\text{largest-}N_E}$
			\ENDIF
			\WHILE {b is true}
			\STATE c $\leftarrow$ c + 1
			\STATE Obtain all $N_E(N-N_E)$ possible selection vectors by applying one swap operation to $\zz_{\text{temp}}$, and compute the corresponding objectives. Let $\zz_{\text{temp-2}}$ be the selection vector among those vectors which yields the minimum objective.
			\IF {$\lvert f(\zz_{\text{temp}})-f(\zz_{\text{temp-2}}) \rvert \leq \mu f(\zz_{\text{temp}})$ \& c $<N_{\text{swap}}^{\text{max}}$}
			\STATE b $\leftarrow$ false, $\zz^{*}_{\text{swap}}\leftarrow  \zz_{\text{temp-2}}$.
			\ELSIF {$c = N_{\text{swap}}^{\text{max}}$}
			\STATE  b $\leftarrow$ false, $\zz^{*}_{\text{swap}}\leftarrow  \zz_{\text{temp-2}}$.
			\ELSE
			\STATE $\zz_{\text{temp}}\leftarrow \zz_{\text{temp-2}}$
			\ENDIF
			\ENDWHILE
		\end{algorithmic}}
	\end{algorithm}
	
\indent{\textit{Remark 3:} It should be noted that the proposed swap algorithm presented in Algorithm~1 reduces to the proposed largest-$N_E$ algorithm if (i) the objective value achieved by the largest-$N_E$ algorithm is sufficiently close to the bound specified by the relaxed algorithm, or (ii) the objective value achieved by the proposed swap algorithm after the first swap operation is the same as that achieved by the largest-$N_E$ algorithm.}

	\subsection{Robust Eavesdropper Selection Problem}\label{sec:Robust}

	In the previous section, it is assumed that the eavesdropper nodes have the perfect knowledge of $\{\lambda_{jk}^{(i)}\}$ (see \eqref{eq:closedform} and \eqref{eq:lambdas}). In this section, we propose a robust eavesdropper selection problem in the presence of imperfect knowledge about the system parameters by introducing some uncertainty in $\{\lambda_{jk}^{(i)}\}$. For simplicity of notation, we assume that $\mathcal{A}_i = \{\yy_1, \yy_2, \ldots, \yy_{N_A}\}$, i.e., all the anchor nodes are connected to the $i$th target position for any $i$. (The proposed approach can easily be extended to scenarios in which this assumption does not hold.)
	
To formulate a robust version of the eavesdropper selection problem, we first define $\bL_E$ as follows:
	\begin{gather}\nonumber
	\bL_E \triangleq \big[\bl_E^{(1)}  ~ \bl_E^{(2)}  \ldots  \bl_E^{(N_T)}\big],
	\end{gather}
where
\begin{gather}\nonumber
\bl_E^{(i)} \triangleq \big[\lambda_{11}^{(i)} ~  \ldots ~ \lambda_{1N}^{(i)}  ~\lambda_{21}^{(i)} ~  \ldots ~  \lambda_{2N}^{(i)} \ldots \lambda_{N_A1}^{(i)} ~   \ldots ~  \lambda_{N_AN}^{(i)} \big]^\intercal.
\end{gather}
	We also introduce the estimated versions of $\bl_E^{(i)}$ as $\blh_E^{(i)}$ for $i = 1, 2, \ldots, N_T$, which are given by
	\begin{gather}
	\blh_E^{(i)} \triangleq  \big[\hat{\lambda}_{11}^{(i)} ~  \ldots ~ \hat{\lambda}_{1N}^{(i)}  ~\hat{\lambda}_{21}^{(i)} ~  \ldots ~  \hat{\lambda}_{2N}^{(i)} \ldots \hat{\lambda}_{N_A1}^{(i)} ~   \ldots ~  \hat{\lambda}_{N_AN}^{(i)} \big]^\intercal
	\end{gather}
	with $\hat{\lambda}_{jk}^{(i)}$ denoting the estimate of $\lambda_{jk}^{(i)}$ for $j= 1,\ldots,N_A$ and $k=1,\ldots,N$. These estimated values represent the imperfect knowledge of the  $\lambda_{jk}^{(i)}$ parameters at the eavesdropper nodes. Let $\bld_E^{(i)}$ denote the error vector that generates the uncertainty; that is,
	\begin{gather}
	\blh_E^{(i)} = \bl_E^{(i)} + \bld_E^{(i)}
	\end{gather}
	with
	\begin{align}
	\bld_E^{(i)}\triangleq  \big[&\Delta\lambda_{11}^{(i)} ~  \ldots ~ \Delta\lambda_{1N}^{(i)}  ~\Delta\lambda_{21}^{(i)} ~  \ldots ~  \Delta\lambda_{2N}^{(i)} \nonumber \\
	& \ldots \Delta\lambda_{N_A1}^{(i)} ~   \ldots ~  \Delta\lambda_{N_AN}^{(i)} \big]^\intercal \label{eq:DelLam_i}
	\end{align}
	for $i = 1,2, \ldots, N_T$. Also, let $\bLd_E$  and $\bLh_E$ be the matrices containing the error vectors and the estimation vectors, respectively, as follows:
	\begin{align}\label{eq:DelLam}
	\bLd_E  & \triangleq \big[\bld_E^{(1)}  ~ \bld_E^{(2)}  \ldots  \bld_E^{(N_T)}\big] \\\label{eq:LamHat}
	\bLh_E  & \triangleq \big[\blh_E^{(1)}  ~ \blh_E^{(2)}  \ldots  \blh_E^{(N_T)}\big].
	\end{align}
	
	In this scenario, the notation for the objective function $f(\bz)$ is modified as $f(\bz, \bL_E)$ to emphasize the dependence on $\bL$ (since $\bLd_E$ becomes another parameter of interest in the presence of uncertainty).
	
	As in \cite{Bertsimas,Bertsimas2,Furkan}, we employ a bounded error model for the uncertainty. In particular, for the eavesdropper selection problem in the presence of parameter uncertainty, the following model is assumed for the error matrix $\bLd_E$:
	\begin{align}\label{eq:robustmodel}
	\bLd_E \in \mathcal{E} \triangleq \big\{\bld^{(i)}\in\mathbb{R}^{N\times N_A}:\,
	{|\Delta\lambda_{jk}^{(i)}|\leq \delta_{jk}^{(i)}}, \forall i, j, k  \big\}
	\end{align}
	where
	$\{\delta_{jk}^{(i)}\}_{i=1,j=1,k=1}^{N_T,N_A,N}$  determine the size of the uncertainty region $\mathcal{E}$ with $\delta_{jk}^{(i)} \geq 0$ for all $i,j,$ and $k$.
	
	The aim is to minimize the worst-case CRLB as in \cite{TCOM2016} and \cite{Furkan}. Therefore, under this setup, the proposed optimization problem can be formulated as
	\begin{subequations}\label{eq:optrobust1}
		\begin{align}
		&\!\min_{\bz}\max_{\bLd_E\in\mathcal{E}}         &\quad& f(\bz, \bL_E) \label{eq:optProb3}\\
		&\text{subject to} &      & \sum_{k=1}^{N} z_k^E  = N_E,\label{eq:constraint31}\\
		&                  &      & 0 \leq z_k^E \leq 1 \text{ for }  k = 1,2, \ldots, N,  \label{eq:constraint32} \\
		&                  &      & \bL_E = \bLh_E- \bLd_E .\label{eq:constraint33}
		\end{align}
	\end{subequations}
	To solve the optimization problem in \eqref{eq:optrobust1}, the following lemma is utilized.
	
	\textit{Lemma 2:} $f(\bz, \bL_E)$ is non-increasing in $\bl^{(i)}$ for all $i = 1,2, \ldots, N_T$.
	
	\textit{Proof:} {See Appendix-D.\hfill $\blacksquare$}

	Let	the value of $\bLd_E$ that maximizes $f(\bz, \bL_E)$ over set $\mathcal{E}$ be denoted as $\bLd_E^{*}$ and let $\{\Delta\lambda_{jk}^{(i),*}\}_{i,j,k}$ represent the elements of $\bLd_E^{*}$ (see \eqref{eq:DelLam_i} and \eqref{eq:DelLam}). Based on Lemma 2, it is obtained that
	\begin{gather}
	\Delta\lambda_{jk}^{(i),*} = \delta_{jk}^{(i)}.
	\end{gather}
	
	Therefore, solving \eqref{eq:optrobust1} is equivalent to solving the following optimization problem:
	\begin{subequations}\label{eq:optrobust1v2}
		\begin{align}
		&\!\min_{\bz}        &\quad& f(\bz, \bLh_E-\bLd_E^{*}) \label{eq:optProb3v2}\\
		&\text{subject to} &      & \sum_{k=1}^{N} z_k^E  = N_E,\label{eq:constraint31v2}\\
		&                  &      & 0 \leq z_k^E \leq 1 \text{ for }  k = 1,2, \ldots, N  .\label{eq:constraint32v2}
		\end{align}
	\end{subequations}
	It is noted that \eqref{eq:optrobust1v2} is in the form of \eqref{eq:opt2}. Thus, the solution approaches discussed for the eavesdropper selection problem in the previous section can also be applied to this problem.

	\section{Jammer Selection Problem}\label{sec:JamSel}

	In this section, we focus on the jammer selection problem under the assumption that there exist only jammer nodes in the environment, i.e., $N_E = 0$. The aim is to choose at most $N_J$ locations from the set $\mathcal{N}$ for jamming purposes so that the target localization performance of the anchor nodes is degraded as much as possible. By using the CRLB of the anchor nodes related to the estimation of target node positions as the performance metric, the jammer selection problem is investigated in the presence and absence of perfect knowledge about the system parameters.

	
\subsection{Problem Formulation}

Let $\zj = [z^{J}_1 ~ \ldots~ z^{J}_N]^{\intercal}$ denote a selection vector defined as
	\begin{equation}\label{eq:zk2}
	z_k^{J}  = \begin{cases}
	1, &\text{if  $\pp_k\in\mathcal{N}_{J}$} \\
	0, &\text{otherwise}
	\end{cases}
	\end{equation}
where $\sum_{k = 1}^{N} z_k^{J} \leq N_J$. Via similar steps to those in \cite{Furkan,part1,TCOM2016}, the EFIM 
related to the positioning of the target node located at $\xx_i$ by the anchor nodes can be obtained as follows:
	\begin{equation}\label{eq:EFIMJammer}
	\tJ_e^{(i)}(\bx_i,\zj) = \sum_{j\in\mathcal{A}_L^{(i)}} \frac{\tilde{\lambda}_{j}^{(i)}}{\tilde{\sigma}_j^2+\sum_{k=1}^{N} z_k^{J}  P_k^{J} \abs{\gamma_{kj}}^2} \bphi_{ij} \bphi_{ij}^{\intercal}
	\end{equation}
In \eqref{eq:EFIMJammer}, $\tilde{\lambda}_{j}^{(i)}$ corresponds to $\lambda_{ij}$ in \cite[Eq.~3]{Furkan}, $\bphi_{ij} = [\cos\varphi_{ij}~\sin\varphi_{ij}]^{\intercal}$, and $\varphi_{ij}$ is the angle from the $i$th target location to $\yy_j$, i.e.,  $\varphi_{ij} = \arctan{\frac{x_{i2}-y_{j2}}{x_{i1}-y_{j1}}}$, where $\yy_{j} \triangleq [y_{j1} ~y_{j2}]^\intercal$.

Based on \eqref{eq:EFIMJammer}, we formulate the proposed jammer selection problem as follows:
	\begin{subequations}\label{eq:optjammer}
		\begin{align}
		&\!\max_{\zj}        &\quad& \sum_{i=1}^{N_T} w_i \, \tr\big\{\big(\tJ_e^{(i)}(\bx_i,\zj)\big)^{-1} \big\} \label{eq:optProbjammer}\\
		&\text{subject to} &      & \sum_{k=1}^{N} z_k^{J} \leq N_J,~\sum_{k=1}^{N} z_k^{J} P_k^{J} \leq P_T ,\label{eq:constraint1jammer}\\
		&                  &      & z_k^{J} \in \{0,1\} \text{ for }  k = 1,2, \ldots, N  \label{eq:constraint4jammer}
		\end{align}
	\end{subequations}
where $P_T$ is total power budget.

For the jammer selection problem in \eqref{eq:optjammer}, the distribution of the target node positions is assumed to be known. {It is also assumed that the anchor node positions, LOS/NLOS conditions for possible target-anchor positions, and $\tilde{\lambda}_{j}^{(i)}$'s are known.} Similar statements to those in Remark 2 can be made for the jammer selection problem, as well. As stated in \cite{GeziciJamPlacement16}, jammer nodes can obtain information about the localization parameters by various means such as  using  cameras  to  learn  the  locations  of anchor  nodes,  performing  prior  measurements  in  the  environment  to  form a  database  for  the  channel  parameters,  and  listening  to  signals  between  anchor and target nodes. {When this information is inaccurate, the robust formulation of the jammer selection problem in Section~\ref{sec:Robust2} can be employed by considering uncertainty in the knowledge of $\tilde{\lambda}_{j}^{(i)}$'s and LOS/NLOS conditions (please also see Remark~6). In addition, the effects of uncertainty in the anchor node positions and in the distribution of the target node position can be evaluated as in Section~\ref{sec:SimNew}.}


\subsection{Theoretical Results}

To simplify the notation, let $\tilde{f}(\zj)$ and $\{g_{ij}(\zj)\}_{i=1,j=1}^{N_T,N_A}$  be defined as
	\begin{align}\label{eq:tildefz}
	\tilde{f}(\zj)&\triangleq \sum_{i=1}^{N_T} w_i \, \tr\big\{\big(\tJ_e^{(i)}(\bx_i,\zj)\big)^{-1} \big\},
\\	\label{eq:gi}
	g_{ij}(\zj)&\triangleq \frac{\tilde{\lambda}_{j}^{(i)}}{\tilde{\sigma}_j^2+\sum_{k=1}^{N} z_k^{J} P_k^{J} \abs{\gamma_{kj}}^2}\,\cdot
	\end{align}
	In the rest of this section, we analyze the convexity and monotonicity properties of $\tilde{f}$ with respect to $\zj$.
	
	\textit{Lemma 3:} $\tilde{f}(\zj)$ is non-decreasing in $\zj$.
	
	\textit{Proof:} {See Appendix-E.\hfill $\blacksquare$}

	\textit{Lemma 4:} $g_{ij}(\zj)$ is a convex function of $\zj$ for any $i,j$.
	
	\textit{Proof:} {See Appendix-F.\hfill $\blacksquare$}

	\textit{Proposition 3:} $\tilde{f}(\zj)$ is a concave function $\zj$.
	
	\textit{Proof:} {See Appendix-G.\hfill $\blacksquare$}

From Lemma 3, we can conclude that if $\mathbf{z}^{*}= [z^{*}_1 ~ z^{*}_2  \ldots z^{*}_N]^\intercal$ is a solution of \eqref{eq:optjammer}, then \eqref{eq:constraint1jammer} must be satisfied with equality, i.e., $\sum_{k=1}^{N} z^{*}_k = N_{J}$ must hold. By relaxing the last constraint in \eqref{eq:constraint4jammer}, the following optimization problem is obtained:\\\vspace{-0.5cm}
		\begin{subequations}\label{eq:optjammer2}
		\begin{align}
		&\!\max_{\zj}        &\quad& \sum_{i=1}^{N_T} w_i \, \tr\big\{\big(\tJ_e^{(i)}(\bx_i,\zj)\big)^{-1} \big\} \label{eq:optProbjammer2}\\
		&\text{subject to} &      & \sum_{k=1}^{N} z_k^{J} = N_J,~\sum_{k=1}^{N} z_k^{J} P_k^{J} \leq P_T, \label{eq:constraint1jammer2}\\
		&                  &      & 0\leq z_k^{J} \leq 1 \text{ for }  k = 1,2, \ldots, N  .\label{eq:constraint4jammer2}
		\end{align}
	\end{subequations}
Since the objective function in \eqref{eq:optProbjammer2} is concave due to Proposition~3 and all the constraints in \eqref{eq:constraint1jammer2} 
and \eqref{eq:constraint4jammer2} are affine, we reach the conclusion that \eqref{eq:optjammer2} is a convex optimization problem. Thus, it can be solved via convex optimization tools for finding its globally optimal solution.

After finding the solution of \eqref{eq:optjammer2}, the largest-$N_J$ algorithm and   {the proposed swap algorithm} can be used for finding the solution of \eqref{eq:optjammer} as in the eavesdropper selection problem. However, in this case, we set the largest $N_J$ components of the solution obtained from \eqref{eq:optjammer2} to one, and while implementing {the proposed swap algorithm}, we check whether there is an increase in the objective function by simply swapping one of the $N_J$ selected positions with one of the $N-N_J$ positions that are not selected.

	
\indent\indent\textit{Remark 4:} For the formulation of \eqref{eq:optjammer2}, it is assumed that the transmit powers of the jammer nodes are given (fixed). If $\{P_k^{J}\}_{k=1}^{N}$ are considered as optimization variables as well, the following problem can be formulated (cf.~\eqref{eq:optjammer2}):	
	\begin{subequations}\label{eq:optjammer3}
		\begin{align}
		&\!\max_{\zj,\tqq}        &~~& \sum_{i=1}^{N_T} w_i \, \tr\Bigg\{\Bigg(\sum_{j\in\mathcal{A}_L^{(i)}} \tilde{g}_{ij}(\tqq) \bphi_{ij} \bphi_{ij}^{\intercal}\Bigg)^{-1} \Bigg\} \label{eq:optProbjammer3}\\
		&\text{subject to} &      & \sum_{k=1}^{N} z^{J}_k = N_J,~\sum_{l=1}^{N} \tq_l \leq P_T , \label{eq:constraint1jammer3}\\
		&                  &      & 0\leq z^{J}_k \leq 1 \text{ for }  k = 1,2, \ldots, N ,\label{eq:constraint3jammer3} \\
		&                  &      & 0\leq \tq_l  \leq  z_l^{J} P_l^{\text{peak}} \text{ for }  l = 1,2, \ldots, N \label{eq:constraint4jammer3}
		\end{align}
	\end{subequations}
where $\tilde{q}_l=z_l^JP_l^J$, $\tqq=[\tilde{q}_1\ldots\tilde{q}_N]^{\intercal}$, $\tilde{g}_{ij}(\tqq)$ is defined as (see \eqref{eq:EFIMJammer})
	\begin{equation}
\tilde{g}_{ij}(\tqq) \triangleq \frac{\tilde{\lambda}_{j}^{(i)}}{\tilde{\sigma}_j^2+\sum_{l=1}^{N} \tq_l \abs{\gamma_{lj}}^2}\,,
	\end{equation}
and $P_{l}^{\text{peak}}$ is the peak power limit for the jammer node located at $\pp_{l}$. It is observed that all the constraints are linear with respect to $\tqq$ and $\zj$ in \eqref{eq:optjammer3}. Furthermore, as a corollary of Proposition 3, one can conclude that the objective function in \eqref{eq:optProbjammer3} is a concave function of $\tqq$. (This holds since there are no assumptions about $\{P_l^{J}\}_{l=1}^{N}$ in Proposition~3 while proving the concavity of the objective function $\tilde{f}(\zj)$ with respect to $\zj$.) Therefore, it is concluded that the optimization problem in \eqref{eq:optjammer3} is convex, as well. This implies that the joint jammer selection and jammer power optimization problem can be solved via the convex problem in \eqref{eq:optjammer3} (after relaxing the selection vector).
	
\indent\indent\textit{Remark 5:} As a special case of \eqref{eq:optjammer3}, it can be shown that the following problem is also convex.	
		\begin{align}\nonumber
		&\!\max_{\tqq}         ~ \sum_{i=1}^{N_T} w_i \, \tr\Bigg\{\Bigg(\sum_{j\in\mathcal{A}_L^{(i)}} \frac{\tilde{\lambda}_{j}^{(i)}}{\tilde{\sigma}_j^2+\sum_{l=1}^{N} \tq_l \abs{\gamma_{lj}}^2} \bphi_{ij} \bphi_{ij}^{\intercal}\Bigg)^{-1} \Bigg\} 
\\\label{eq:optjammer4}
		& 	\text{s.t.}	   	  ~ \sum_{j=1}^{N} \tq_j \leq P_T ,~ 
0\leq \tq_l  \leq   P_l^{peak} \text{ for }  l = 1,2, \ldots, N  . 
		\end{align}
It is noted that this problem is in the same form as the problem discussed in \cite[Eq.~9]{Furkan}. In \cite{Furkan}, the convexity of this problem is not taken into account. Instead, a series of geometric programming approximations are proposed in order to solve the optimization problem. Since the problem \cite[Eq.~9]{Furkan} is in fact convex, it can also be solved via convex optimization tools.
	
\vspace{-0.2cm}
	
	\subsection{Robust Jammer Selection Problem} \label{sec:Robust2}

	In the previous section, the jammer nodes are assumed to have the perfect knowledge of $\{\tilde{\lambda}_{j}^{(i)}\}_{i = 1, j = 1}^{N_T,N_A}$ in \eqref{eq:EFIMJammer}. Similar to Section~\ref{sec:Robust}, some uncertainty in $\{\tilde{\lambda}_{j}^{(i)}\}_{i = 1, j = 1}^{N_T,N_A}$ is introduced for a robust formulation. (No uncertainty is considered for $|\gamma_{kj}|^2$'s in \eqref{eq:EFIMJammer} since they mainly depend on the known positions of the jammer and anchor nodes.) For simplicity, it is assumed that $\mathcal{A}_i = \{\yy_1, \yy_2, \ldots, \yy_{N_A}\}$, i.e., all the anchor nodes are connected to the $i$th target position for any $i$.

To formulate the robust jammer selection problem, we first define $\bL_J \triangleq [\bl^{(1)}_J  ~ \bl^{(2)}_J  \ldots  \bl^{(N_T)}_J]$, where $\bl_J^{(i)} \triangleq [\tilde{\lambda}_1^{(i)} ~ \tilde{\lambda}_2^{(i)}  \ldots  \tilde{\lambda}_{N_A}^{(i)}]^\intercal$ for $i=1,\ldots,N_T$.
	The estimated versions of $\bl_J^{(i)}$ are defined as $\blh_J^{(i)}$ for $i = 1, 2, \ldots, N_T$, where $\blh_J^{(i)}$ denotes the estimate of $\bl_J^{(i)}$. Let $\bld_J^{(i)}$ represent the error vector that generates uncertainty, that is, $\blh_J^{(i)} = \bl_J^{(i)} + \bld_J^{(i)}$
	with
	\begin{equation}\label{eq:DelLam_i2}
	\bld_J^{(i)}\triangleq [\Delta\tilde{\lambda}_1^{(i)} ~ \Delta\tilde{\lambda}_2^{(i)}  \ldots  \Delta\tilde{\lambda}_{N_A}^{(i)}]^\intercal
	\end{equation}
	for $i=1, 2, \ldots, N_T$. Also, $\bLd_J$ and $\bLh_J$ are defined as
	\begin{align}\label{eq:DelLam2}
	\bLd_J  & \triangleq [\bld_J^{(1)}  ~ \bld_J^{(2)}  \ldots  \bld_J^{(N_T)}], \\\label{eq:LamHat2}
	\bLh_J  & \triangleq [\blh_J^{(1)}  ~ \blh_J^{(2)}  \ldots  \blh_J^{(N_T)}].
	\end{align}
	
	In this scenario, the notation for the objective function $\tilde{f}(\zj)$ is modified as $\tilde{f}(\zj,\bL_J)$ in order to emphasize the dependence on $\bL_J$. We use the same bounded error model as in Section~\ref{sec:Robust} for the error matrix $\bLd_J$:	
	\begin{align}\nonumber
	\bLd_J \in \tilde{\mathcal{E}} & \triangleq \big\{\bld^{(i)}_J\in\mathbb{R}^{N_A}:\,
	{|\Delta\tilde{\lambda}_j^{(i)}|\leq \tilde{\delta}_{j}^{(i)}},  \\ &\forall i = 1,2,  \ldots, N_T
	{ \text{ and } \forall j = 1,2,\ldots, N_A} \big\} \label{eq:robustmodel2}
	\end{align}
	where $\{\tilde{\delta}_{j}^{(i)}\}_{i=1,j=1}^{N_T,N_A}$  determine the size of the uncertainty region $\tilde{\mathcal{E}}$  with $ \tilde{\delta}_{j}^{(i)} \geq 0$  for all $i = 1, 2, \ldots,N_T$ and $j = 1,2,\ldots, N_A$.
	
	The aim is to maximize the minimum CRLB that can be achieved in $\tilde{\mathcal{E}}$. Therefore, under this setup, the proposed optimization problem can be formulated as
	\begin{subequations}\label{eq:optrobust2}
		\begin{align}
		&\!\max_{\zj}\min_{\bLd_J\in\tilde{\mathcal{E}}}         &\quad& \tilde{f}(\zj, \bL_J) \label{eq:optProb4}\\
		&\text{subject to} &      & \sum_{k=1}^{N} z_k^{J}  = N_J,~\sum_{k=1}^{N} z_k^{J}  P_k^{J} \leq P_T , \label{eq:constraint41}\\
		&                  &      & 0 \leq z_k^{J} \leq 1 \text{ for }  k = 1,2, \ldots, N,  \label{eq:constraint42} \\
		&                  &      & \bL_J = \bLh_J- \bLd_J.\label{eq:constraint45}
		\end{align}
	\end{subequations}
	To solve the optimization problem in \eqref{eq:optrobust2}, the following lemma is utilized.
	
	\textit{Lemma 5:} $\tilde{f}(\zj,\bL_J)$ is non-increasing in $\bl_J^{(i)}$ for all $i = 1,2, \ldots, N_T$.
	
	\textit{Proof:} {See Appendix-H.\hfill $\blacksquare$}

	Let 
	the value of $\bLd_J$ that minimizes $\tilde{f}(\zj,\bL_J)$ over set $\tilde{\mathcal{E}}$ be denoted as $\bLd_J^{*}$ and let $\{\Delta\tilde{\lambda}_j^{(i),*}\}_{i,j}$ represent the elements of $\bLd_J^{*}$ (see \eqref{eq:DelLam_i2} and \eqref{eq:DelLam2}). Based on Lemma 5, it is obtained that ${\Delta\tilde{\lambda}_j^{(i),*} = -\tilde{\delta}_j^{(i)}}$.
Therefore, solving \eqref{eq:optrobust2} is equivalent to solving the following optimization problem:
\begin{subequations}\label{eq:optrobust2v2}
	\begin{align}
	&\!\max_{\zj}        &\quad& \tilde{f}(\zj, \bLh_J-\bLd_J^{*}) \label{eq:optProb4v2}\\
	&\text{subject to} &      & \sum_{k=1}^{N} z_k^{J}  = N_J,~\sum_{k=1}^{N} z_k^{J} P_k^{J} \leq P_T,\label{eq:constraint41v2}\\
	&                  &      & 0 \leq z_k^{J} \leq 1 \text{ for }  k = 1,2, \ldots, N.  \label{eq:constraint42v2}
	\end{align}
\end{subequations}
This problem is exactly in the same form as the problem in \eqref{eq:optjammer2}, hence, it is a convex optimization problem. Therefore, the solution methods proposed for the jammer selection problem can also be used for the problem in \eqref{eq:optrobust2v2}.

\textit{Remark~6:} The imperfect knowledge of LOS/NLOS conditions can be incorporated into the $\lambda_{jk}^{(i)}$ and $\tilde{\lambda}_j^{(i)}$ parameters in Sections~\ref{sec:EavSel} and \ref{sec:JamSel}. (In the case of a NLOS link, the corresponding $\lambda_{jk}^{(i)}$ and $\tilde{\lambda}_j^{(i)}$ parameters become zero; i.e., no position related information is gathered from that link.) Hence, the cases with imperfect knowledge of LOS/NLOS conditions can be treated in the robust eavesdropper and jammer selection approaches in Sections~\ref{sec:Robust} and \ref{sec:Robust2}.

\section{Joint Eavesdropper and Jammer Selection}\label{sec:JointSel}

In this section, we consider the eavesdropper and jammer selection problems jointly
and 
place jammer and eavesdropper nodes by considering both the localization performance of the anchor nodes (which is to be degraded) and the accuracy of the eavesdropper nodes for estimating the location of the target node (which is to be enhanced). In this part, it is assumed that the jammer nodes do not cause any interference at the eavesdropper nodes; e.g., by using directional antennas towards the anchor nodes. {In addition, we make the same assumptions as in the eavesdropper selection problem and the jammer selection problem.}



Based on the selection vectors $\bz$ and $\zj$, the joint eavesdropper and jammer selection problem can be formulated as
\begin{subequations}\label{eq:optjoint}
	\begin{align}
	&\!\max_{\zj,\bz}        &\quad& \tilde{f}(\zj) \\
	& \text{subject to}&     & f(\bz) \leq \rho, ~\sum_{k=1}^{N} z_k^{E}  = N_E, \\
	& 				   &      & \sum_{k=1}^{N} z_k^{J}  = N_J,~\sum_{k=1}^{N} z_k^{J} P_k^{J} \leq P_T,\\
	&				   &	  & z_{k}^{E} \in \{0,1\}  \text{ for }  k = 1,2, \ldots, N  \label{eq:z_eav} \\
	&				   &	  & z_{k}^{J} \in \{0,1\}  \text{ for }  k = 1,2, \ldots, N  \label{eq:z_jam} \\
	&				   &	  & z_{k}^{E}z_{k}^{J} = 0   \text{ for }  k = 1,2, \ldots, N  \label{eq:valid}
	\end{align}
\end{subequations}
where $\tilde{f}(\zj)$ is as in \eqref{eq:tildefz}, $f(\bz)$ is given by \eqref{eq:fz}, and $\rho$ is a given accuracy threshold related to eavesdropping.
The last constraint \eqref{eq:valid} guarantees that a node can be selected either as an eavesdropper or as a jammer. By relaxing the constraints in \eqref{eq:z_eav}  and \eqref{eq:z_jam}, and modifying \eqref{eq:valid}, we  obtain the following optimization problem:
\begin{subequations}\label{eq:optjoint2}
	\begin{align}
	&\!\max_{\zj,\bz}        &\quad& \tilde{f}(\zj) \\
	& \text{subject to}&     & f(\bz) \leq \rho ,~\sum_{k=1}^{N} z_k^{E}  = N_E,\\
	& 				   &      & \sum_{k=1}^{N} z_k^{J}  = N_J,~\sum_{k=1}^{N} z_k^{J} P_k^{J} \leq P_T,\\
	&				   &	  & 0\leq z_{k}^{E} \leq 1 \text{ for }  k = 1,2, \ldots, N  \\
	&				   &	  & 0\leq z_{k}^{J} \leq 1  \text{ for }  k = 1,2, \ldots, N  \\
	&				   &	  & 0\leq z_{k}^{E}+z_{k}^{J} \leq 1 \text{ for }  k = 1,2, \ldots, N.
	\end{align}
\end{subequations}
As consequences of Proposition 2 and 3, it is noted that the optimization problem \eqref{eq:optjoint2} is a convex optimization problem.

{The selection of $\rho$ depends on the requirements in a given scenario. For instance, if learning the positions of the target nodes is more important than jamming the localization network, $\rho$ should be small.} Alternatively, one can try to minimize $f(\bz)$ while keeping $\tilde{f}(\zj)$ above a certain threshold. From Proposition 2 and 3, it can be argued that the resulting problem would also be convex. Hence, by using convex optimization tools, the solution of \eqref{eq:optjoint2} or its alternative version can be obtained. Then, starting from that solution, the largest-$N_J$ (or, largest-$N_E$) and swap algorithms can be used to obtain the solution of \eqref{eq:optjoint} or its alternative version.


\section{Simulation Results}\label{sec:Simu}

\begin{figure}
	\includegraphics[width=0.9\linewidth]{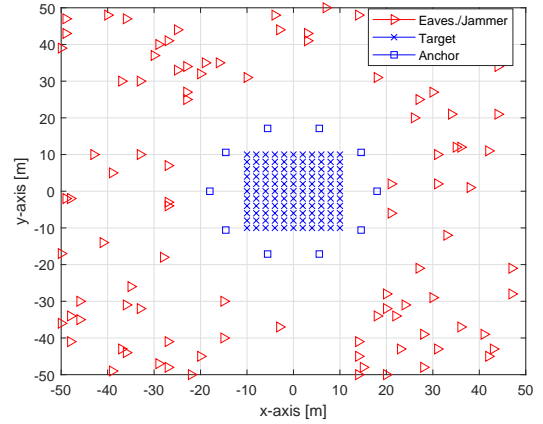}
	\centering
	\caption{{Illustration of the wireless source localization network.}}
	\label{fig:Network}
\end{figure}

In this section, simulations are conducted to investigate the performance of the proposed approaches. We consider a wireless source localization network, in which the target node is located at one of the {121} possible positions with equal probabilities (i.e., {1/121}). In particular,   {the set of possible target positions is given by $\{\xx_i\}_{i=1}^{121} =\{[2m,2n]\mid\ -5\leq m, n\leq 5, m, n \in \mathbb{Z}\}$ meters.}  Also, there are 10 anchor nodes at locations {$\{\yy_j\}_{j=1}^{10} = $$\{[18\cos(\psi_j), 18\sin(\psi_j)] \mid \psi_j = 2\pi (j-1)/10, j = 1, 2, \ldots, 10\}$ meters.} In addition, there exists {100} possible positions for the eavesdropper and jammer nodes {which are selected uniformly from the region $\mathcal{R} = \left([20,50]\times[-50,50]\right)\cup$ $\left([-50,-20]\times[-50,50]\right)\cup\left([-20,20]\times[-50,-30]\right)\cup$$\left([-20,20]\times[30,50]\right)$ meters. Such a region is selected in order to keep eavesdropper/jammer nodes away from the localization network by considering a practical application scenario as in Section~\ref{sec:Moti}. Fig.~\ref{fig:Network} illustrates the positions of the target and anchor nodes, as well as the possible positions for the eavesdropper and jammer nodes.}
	
In the simulations, we consider the eavesdropper selection problem, the jammer selection problem, and the joint eavesdropper and jammer selection problem, given by \eqref{eq:opt}, \eqref{eq:optjammer}, and \eqref{eq:optjoint}, respectively. For the problem in \eqref{eq:optjoint}, we assume that $N_E+N_J = N$. In other words, we have $\bz_k = 1-\zj_k$ for any $k$, for the joint eavesdropper and jammer selection problem.

The following algorithms are investigated for performance comparisons:
\begin{itemize}[leftmargin=*]
	\item \textit{Relaxed {Algorithm}}: The relaxed versions of \eqref{eq:opt}, \eqref{eq:optjammer}, and \eqref{eq:optjoint} (see \eqref{eq:opt2}, \eqref{eq:optjammer2}, and \eqref{eq:optjoint2}) are solved via 
the \text{fmincon}$(\cdot)$ command of MATLAB {by using the interior point algorithm, which has polynomial-time complexity in the worst case, and is very fast in practice}. The solution of \eqref{eq:opt2} provides a lower bound for \eqref{eq:opt}, whereas the solutions of \eqref{eq:optjammer2} and \eqref{eq:optjoint} provide upper bounds for \eqref{eq:optjammer} and \eqref{eq:optjoint}, respectively.
	\item \textit{Largest-$N_E$ {Algorithm}}: We set the largest $N_E$ components of the solution of \eqref{eq:opt2} to one and the others to zero, and we evaluate the performance of this resulting selection vector using the expression in \eqref{eq:closedform}. 
	\item \textit{Largest-$N_J$ {Algorithm}}: In this algorithm, we set the largest $N_J$ components of the solution of \eqref{eq:optjammer2} to one, and the others to zero. For the problem in \eqref{eq:optjoint}, if the relaxed solution pair obtained from \eqref{eq:optjoint2} is denoted as $(\bz_{\text{relaxed}},\zj_{\text{relaxed}})$, we simply set the largest $N_J$ components of $\zj_{\text{relaxed}}$ to one and the others to zero. The resulting vector is denoted as $\zj_{\text{largest}}$, and $\bz_{\text{largest}}$ is defined as $\boldsymbol{1}-\zj_{\text{largest}}$, where $\boldsymbol{1}$ is the vector of ones. (The solution pair $(\bz_{\text{largest}},\zj_{\text{largest}})$ may not be feasible for \eqref{eq:optjoint} unless the threshold value, $\rho$, is sufficiently large.)
	\item \textit {{Proposed Swap Algorithm:}} In this algorithm, {we start} from the solutions obtained from the largest-$N_E$ or the largest-$N_J$ algorithms. The swap operation is performed as explained in Sections~\ref{sec:EavSel} and \ref{sec:JamSel}{, the details of which are given in Algorithm 1. In all the simulations, $\mu$ in Algorithm 1 is selected as $0.01$.} {During one swap operation, the number of objective function evaluations is given by $N_E(N-N_E)$. In other words, the total number of objective evaluations is upper bounded by $N_{\text{swap}}^{\text{max}}(N-N_E)N_E$ (similarly for the jammer selection problem).}
	{\item \textit {Swap Algorithm with Random Initialization:} This algorithm is considered for comparison purposes similar to the local optimization algorithm in \cite{Berkay2020}. In this algorithm, we use the proposed swap algorithm (Algorithm~1) with arbitrarily generated initial selection vectors (inputs) for the eavesdropper selection problem or the jammer selection problem. While generating the random initial vectors, we randomly choose $N_E$ or $N_J$ positions from $N$ possible eavesdropper/jammer positions by using the \textit{randperm($N$,$N_E$)} or \textit{randperm($N$,$N_J$)} command of MATLAB with different seeds.}
\end{itemize}

For the eavesdropper selection problem, we assume that $\sigma_k^2 = \sigma^2$ for each $k$. Moreover, $\alpha_{ijk}^{(E,1)}$ and $\chi_{jk}^{(i)}$ are modeled as $\big|\alpha_{ijk}^{(E,1)}\big|^2 = \norm{\xx_i-\pp_k}^{-2}$ and  $\chi_{jk}^{(i)} = 0$. Hence, $\lambda_{jk}^{(i)}$ is expressed as   $\lambda_{jk}^{(i)} = 4\pi \beta_{ij}^2 E_{ij} /(c^2 \norm{\xx_i-\pp_k}^2 \sigma^2)$, where $E_{ij} = \int_{-\infty}^{\infty} |S_{ij}(f)|^2\,df$ is the energy of the signal $s_{ij}(t)$ (see Proposition~1). Then, the signal parameters are selected such that $\lambda_{jk}^{(i)}$ is given by $\lambda_{jk}^{(i)} = 1/(\norm{\xx_i-\pp_k}^2 \sigma^2)$ \cite{JammingWCL}.

For the jammer selection problem, it is assumed that $\tilde{\sigma}_j^2 = \tilde{\sigma}^2$ for each $j$, $\tilde{\lambda}_{j}^{(i)} = 1/(\norm{\xx_i-\yy_j}^2 )$, and $\abs{\gamma_{kj}}^2 =  \norm{\pp_k-\yy_j}^{-2}$. Regarding the transmit powers of the jammer nodes, $P_k^{J} = {10}$ for each $k$ and $P_T$ is selected as ${10N}$, i.e., the constraint given by $\sum_{k} z_k^{J} P_k^{J}\leq P_T$ becomes ineffective.

{In order to perform simulations considering the shadowing effect, $\tilde{\lambda}_j^{(i)}$'s and $\lambda_{jk}^{(i)}$'s are multiplied with log-normal random variables with mean parameter $-2$ and variance parameter $1$. Similarly, $\abs{\gamma_{kj}}^2$'s are multiplied with log-normal random variables with mean parameter $-2$ and variance parameter $2$.}

{In the simulations, for each problem, the square roots of the objectives are plotted, i.e., the average and the worst-case CRLB values are presented in terms of meters. The simulations are performed on an Intel Core i7 4.0 GHz PC with 16 GB of physical memory using MATLAB R2020b on a Windows 10 operating system.}

\subsection{Simulation Results with Perfect Knowledge of Parameters}\label{sec:SimNonRobust}

\begin{figure}
	\includegraphics[width=0.9\linewidth]{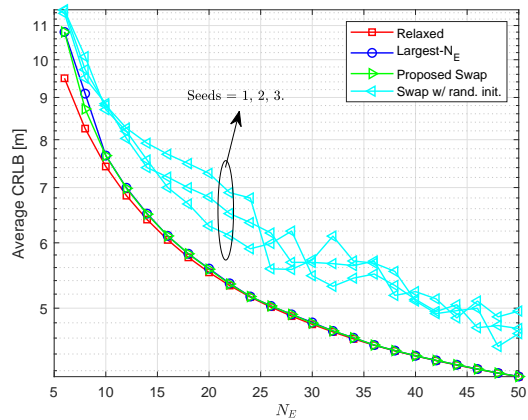}
	\centering
	\caption{{Average CRLB versus $N_E$ when $\sigma^2 = 0.1$, $N_{\text{swap}}^{\text{max}} = 5$, and the seeds of the random initial selection vectors are $1, 2, 3$ for the eavesdropper selection problem.}}
	\label{fig:EavCRLBvsNE}
\end{figure}

\begin{figure}
	\includegraphics[width=0.9\linewidth]{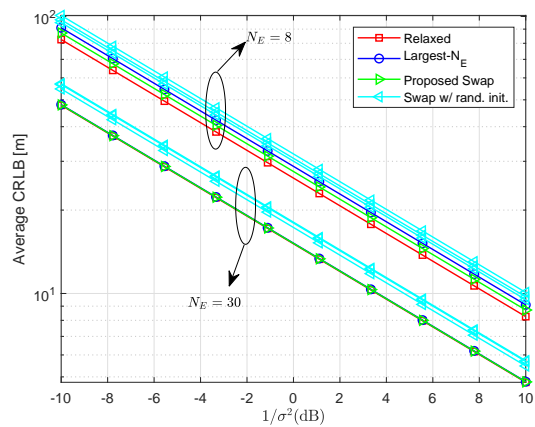}
	\centering
	\caption{{Average CRLB versus $1/\sigma^2$ when $N_E = 8$, $N_E= 30$, $N_{\text{swap}}^{\text{max}} = 5$, and the seeds of the random initial selection vectors are $1, 2, 3$ for the eavesdropper selection problem.}}
	\label{fig:EavCRLBvsSNR}
\end{figure}

In Fig.~\ref{fig:EavCRLBvsNE}, the eavesdropper selection problem is considered and the average CRLB performance of each algorithm is plotted versus $N_E$ for the noise level {$\sigma^2 = 0.1$ and $N_{\text{swap}}^{\text{max}} = 5$}. For the same setting, Fig.~\ref{fig:EavCRLBvsSNR} presents the average CRLB performance of each algorithm versus $1/\sigma^2$ for {$N_{\text{swap}}^{\text{max}} = 5$ and two different levels of $N_E$'s: $N_E = 8$ and $N_E = 30$}. From Figs.~\ref{fig:EavCRLBvsNE} and \ref{fig:EavCRLBvsSNR}, it is observed that the solution of the relaxed problem provides a performance lower bound, as expected, and {the largest-$N_E$ algorithm and the proposed swap algorithm} perform very similarly in this scenario.  {On the other hand, when the swap algorithm is executed based on three different random initial selection vectors (with seeds $1$, $2$, and $3$), significant performance degradation is observed in comparison with the other algorithms. This implies that solving the relaxed problem and then obtaining the solution of the largest-$N_E$ algorithm or the proposed swap algorithm is critical in achieving high localization accuracy.}

As $\sigma_k^2 = \sigma^2$ for all $k = 1,2, \ldots, N$, it is noted that by changing  $\sigma^2$, we in fact scale all $\lambda_{jk}^{(i)}$'s with the same factor. Therefore, by Remark~1, it is concluded that the objective function is also scaled, as can be observed from Fig.~\ref{fig:EavCRLBvsSNR}. Moreover, from Remark~1, it is known that the solution of {the optimal eavesdropper selection problem (hence, that of the largest-$N_E$ algorithm) remains} the same for all $\sigma^2$'s when $N_E$ is fixed. For instance, when {there are $8$ eavesdroppers in the network, the $24, 33, 38, 39, 51, 77, 88, 92$th components of $\bz_{\text{largest}}$ are equal to $1$ for both $\sigma^2 = 0.1$ and $\sigma^2 = 10$}.

{The average CRLB performance and run time of each algorithm are evaluated versus $N_{\text{swap}}^{\text{max}}$
for} $\sigma^2 = 0.1$ and $N_E = 15$. {(The figures are not presented due to the space constraint.) The results indicate that
it requires around $13$ swap operations for the swap algorithm with random initialization (with seed $1$) to converge to the performance of the proposed swap algorithm. Namely, the average CRLB of the swap algorithm with random initialization is $11.4$ m at $N_{\text{swap}}^{\text{max}}=1$ and reduces to that of the proposed swap algorithm (i.e., $6.27$ m) at $N_{\text{swap}}^{\text{max}}=13$.}
On the other hand, the starting point obtained by the proposed largest-$N_E$ algorithm {($6.27$ m)} is not improved by the proposed swap algorithm, i.e., the largest-$N_E$ algorithm provides the best selection vector in this scenario (please see Algorithm~1). When the corresponding run times in 
are compared, the benefits of the proposed swap and largest-$N_E$ algorithms are observed. {While the run time of the proposed swap algorithm is $0.9$ sec. for each $N_{\text{swap}}^{\text{max}}$, that of the swap algorithm with random initialization is $10.11$ sec. for $N_{\text{swap}}^{\text{max}}=13$.} Thanks to the relaxed algorithm, the proposed swap algorithm starts with a selection vector which is very close to the optimal selection vector; hence, it obtains the solution quickly. On the other hand, with random initial selection vectors, high localization accuracy cannot be obtained without performing a time-consuming search based on swap operations.

\begin{figure}
	\includegraphics[width=0.9\linewidth]{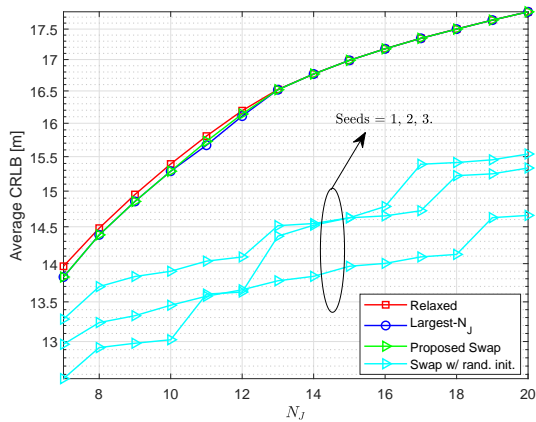}
	\centering
	\caption{{Average CRLB versus $N_J$ when $\tilde{\sigma}^2 = 0.1$, $N_{\text{swap}}^{\text{max}} = 5$, and the seeds of the random initial selection vectors are $1, 2, 3$ for the jammer selection problem.}}
	\label{fig:JamCRLBvsNJ}
\end{figure}
\begin{figure}
	\includegraphics[width=0.9\linewidth]{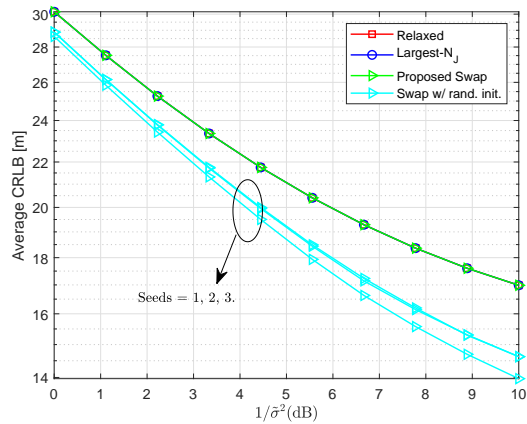}
	\centering
	\caption{{Average CRLB versus $1/\sigma^2$ when $N_J= 15$,  $N_{\text{swap}}^{\text{max}} = 5$, and the seeds of the random initial selection vectors are $1, 2, 3$ for the jammer selection problem.}}
	\label{fig:JamCRLBvsSNR}
\end{figure}

In Fig.~\ref{fig:JamCRLBvsNJ}, the jammer selection problem is considered and the average CRLB performance of each algorithm is plotted versus $N_J$ {for the noise level $\tilde{\sigma}^2 = 0.1$.} For the same setting, Fig.~\ref{fig:JamCRLBvsSNR} presents the average CRLB performance of each algorithm versus $1/\tilde{\sigma}^2$ for ${N_J = 15}$. From Figs.~\ref{fig:JamCRLBvsNJ} and \ref{fig:JamCRLBvsSNR}, it is observed that the solution of the relaxed problem provides a performance upper bound, as expected, and {the proposed largest-$N_J$ algorithm and the proposed swap algorithm perform similarly. However, when the proposed swap algorithm is implemented based on three different random initial jammer selection vectors (instead of the solution of the largest-$N_J$ algorithm), the obtained CRLB values reduce significantly. This indicates the advantage of the proposed approaches over the swap algorithm with random initialization.}

{The CRLB performance and run time of each algorithm are evaluated versus $N_{\text{swap}}^{\text{max}}$ for $\tilde{\sigma}^2 = 0.1$ and $N_J = 15$. (The figures are not presented due to the space constraint.) The results indicate that after around $13$ swap operations, the average CRLB of the swap algorithm with random initialization (which is initially $10.67$ m.) converges that of the proposed swap algorithm (i.e., $16.98$ m). (In this scenario,} the starting point obtained by the proposed largest-$N_J$ algorithm already corresponds to the best selection vector.{) While the run time of the proposed swap algorithm is $0.2$ sec., it takes around $5.24$ sec. for the swap algorithm with random initialization (with seed $1$) to converge to the proposed swap algorithm. Hence,} the proposed swap and largest-$N_J$ algorithms have significantly lower execution times than the swap algorithm with random initialization {considering the same CRLB performance}. This indicates that the proposed approach of solving the relaxed algorithm and using its solution as a basis for the largest-$N_J$ and the swap algorithms provides significant benefits in obtaining the solution of the optimal jammer selection problem. In other words, the swap algorithm cannot achieve close to optimal performance in a short amount of time by starting from a random selection vector.


\begin{figure}
	\centering
	\includegraphics[width=0.9\linewidth]{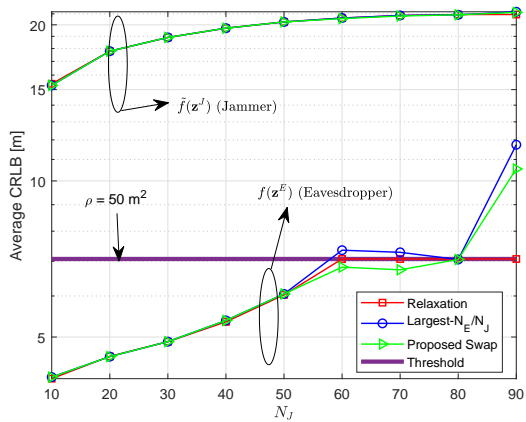}
	\vspace{-0.25cm}
	\caption{{Average CRLB versus $N_J$ when $\tilde{\sigma}^2 =  \sigma^2 = 0.1$ and ${\rho = 50}$ for the joint eavesdropper and jammer selection problem.}}
	\vspace{-0.5cm}

	\label{fig:Joint}
\end{figure}

In Fig.~\ref{fig:Joint}, the joint eavesdropper and jammer selection problem is investigated, and the average CRLB performances corresponding to the objective functions $f(\bz)$ and $\tilde{f}(\zj)$ are plotted for each algorithm when {$\rho=50$} (see \eqref{eq:optjoint} and \eqref{eq:optjoint2}).
It is calculated that for ${N_J = 60, 70, 90}$, or equivalently ${N_E = 40, 30, 10}$, the solution of the largest-$N_E$ algorithm is not a feasible solution for \eqref{eq:optjoint}. {For example, when $N_J = 60$}, the average CRLB for the largest-$N_E$ algorithm is {$54.06$}, which is higher than {$\rho = 50$}. Also, even though the {solutions} of the largest-$N_E$ algorithm are infeasible for {$N_J = 60, 70$}, starting from {these solutions, via the proposed swap algorithm, it is possible obtain feasible selection vectors without reducing the value of $\tilde{f}(\zj)$. However, when $N_J = 90$, via the proposed swap algorithm, it is not possible to obtain a feasible selection vector.} Moreover, a decrease is observed in the optimal value of {$f(\bz)$} from {$N_J = 60$ to $N_J = 70$, or equivalently from $N_E = 40$ to $N_E = 30$.} In other words, it is not possible to claim any monotonic behavior in {$f(\bz)$} with respect to {$N_E$} due to the constraint given by $f(\bz)\leq \rho$ for the problem in \eqref{eq:optjoint}. Furthermore, the relaxed problem does not necessarily provide a lower bound on $f(\bz)$ as noted from the results at {$N_J =60$ and $N_J=70$ (equivalently, $N_E =40$ and $N_E=30$)}.

\subsection{{Effects of Uncertainty in Knowledge of Target and/or Anchor Locations}}\label{sec:SimNew}

{In this part, we introduce some uncertainty to the knowledge related to the locations of the anchor and target nodes, and obtain the optimal selection strategies (using the relaxed formulations) for the cases of imperfect and perfect knowledge. Then, we apply the largest-$N_E/N_J$ and proposed swap algorithms and evaluate their performance based on the actual system parameters.}

For the eavesdropper selection problem,
we consider a scenario in which the eavesdropper nodes do not know the probability distribution of the target node location perfectly. (The knowledge of anchor node locations is not required for the eavesdropper selection problem.) In particular, for $i = 1, 2, \ldots, N_T$, the actual distribution of the target node location is given by
	\begin{equation}
		 \tilde{w}_i = A \exp\left(-\frac{(x_{i1}-x_{01})^2}{2\nu^2}-\frac{(x_{i2}-x_{02})^2}{2\nu^2}\right)
		\label{eq:pdist}
\end{equation}
where $\tilde{w}_i =\Pr\{\text{Target node is located at } \xx_i \}$, $\xx_i = [x_{i1} \, x_{i2}]^{\intercal}$, $\bar{x} = [x_{01}\, x_{02}]^{\intercal}$ is the mean of the target node location, and $A$ is a normalization constant such that
	$\sum_{i=1}^{N_T} A \exp\left(-\frac{(x_{i1}-x_{01})^2}{2\nu^2}-\frac{(x_{i2}-x_{02})^2}{2\nu^2}\right) = 1$.
On the other hand, the eavesdropper nodes assume that $\Pr\{\text{Target node is located at } \xx_i \} = 1/N_T$ for $i = 1, 2, \ldots, N_T$. It is noted that as $\nu$ tends to infinity, $\tilde{w}_i$ approaches to $1/N_T$ for each $i$. In other words, as $\nu$ increases, the mismatch between the true distribution and the assumed one decreases. On the other extreme, when $\nu$ goes to zero, the target node is located at $\bar{x}$ with probability one; hence, the uniform distribution assumption becomes quite inaccurate.

\begin{figure}
	\includegraphics[width=0.9\linewidth]{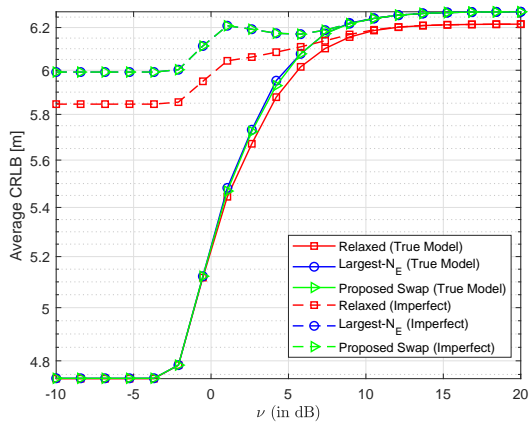}
	\centering
	\caption{{Average CRLB versus $\nu$ (in dB) when $N_E = 15$, $\sigma^2 = 0.1$, and $\mu = 0.01$.}}
	\label{fig:CRLBvsNu_N_E=15}
\end{figure}

In the simulations, we assume that $x_{01} = x_{02} = 0$ and $N_{\text{swap}}^{\text{max}} = 5$. In Fig.~\ref{fig:CRLBvsNu_N_E=15}, the average CRLB performance of each algorithm is plotted versus $\nu$ in dB (i.e., $10\log_{10}\nu$) for $N_E = 15$, $\sigma^2 = 0.1$, and $\mu = 0.01$. It is observed that as long as the information about the distribution of the target node location is not very inaccurate (i.e., $\nu$ is not very small), the proposed approach does not have a significant performance loss. Also, as the mismatch between the true distribution and the assumed one decreases (i.e., as $\nu$ increases), the proposed swap algorithm performs very similarly for both the true model and the assumed one.

\begin{figure}
	\includegraphics[width=0.9\linewidth]{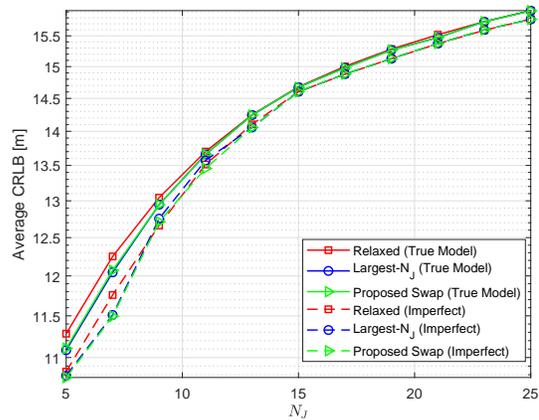}
	\centering
	\caption{{Average CRLB versus $N_J$ when   $\tilde{\sigma}^2 = 0.1$, $\nu = 1$, $r = 1$, and $\mu = 0.01$.}}
	\label{fig:JamCRLBvsNJ_r=1_nu_1}
\end{figure}

For the jammer selection problem, we assume that the jammer nodes do not know the locations of the anchor nodes perfectly. It is assumed that for any $\yy_j =[y_{j1} \, y_{j2}]^{\intercal} $, the jammer nodes have the knowledge of an erroneous version of $\yy_j$. Let $\tilde{\yy}_j$ be the assumed location of the $j$th anchor node by the jammer nodes. We model that $\tilde{\yy}_j$ is uniformly chosen from a set $\{\yy \mid \yy = [y_1 \, y_2]^{\intercal}, \abs{y_1-y_{j1}}\leq r \, \& \, \abs{y_2-y_{j2}}\leq r  \}$. In Fig.~\ref{fig:JamCRLBvsNJ_r=1_nu_1}, when $\tilde{\sigma}^2 = 0.1$, $\nu = 1$, $r=1$, and $\mu = 0.01$, the average CRLB performance of each algorithm is plotted versus $N_J$. It is observed that the proposed swap algorithm is quite robust to errors in the knowledge of anchor and target node locations. Even though the anchor node locations and the distribution of the target node location are not known perfectly, as $N_J$ increases, the proposed swap algorithm performs very similarly for both the true model and the assumed one.

\subsection{{Simulation Results for Robust Approaches}}\label{sec:SimRobust}

In this part, the robust eavesdropper selection problem in Section~\ref{sec:Robust} and the robust jammer selection problem in Section~\ref{sec:Robust2} are considered. The worst-case CRLB performances of the algorithms are compared for both the robust and non-robust approaches. In the robust approach, the problems given by \eqref{eq:optrobust1v2} and \eqref{eq:optrobust2v2} are considered for the robust eavesdropper and the robust jammer selection problems, respectively. However, in the non-robust case, the following optimization problems are considered: $\min_{\bz}f(\bz, \bLh_E)$ subject to $\sum_{k=1}^{N} z_k^E  = N_E$, $0 \leq z_k^E \leq 1$ for $k = 1,2, \ldots, N$, which is the non-robust version of the eavesdropper selection problem, and $\max_{\zj}\tilde{f}(\zj, \bLh_J)$ subject to $\sum_{k=1}^{N} z_k^{J}  = N_J$, $0 \leq z_k^{J} \leq 1$ for $k = 1,2, \ldots, N$, which is the non-robust version of the jammer selection problem.

For the eavesdropper selection problem, both the robust and non-robust approaches are considered, and two different selection vectors denoted as $\bz_{R}$ and $\bz_{NR}$ (corresponding to robust and non-robust, respectively) are obtained for each algorithm. Then, for $\bz_{R}$ and $\bz_{NR}$, the corresponding worst-case CRLBs are computed, which are given by  $f(\bz_{R}, \bLh_E-\bLd_E^{*})$ and $f(\bz_{NR},\bLh_E-\bLd_E^{*})$, respectively. Similarly, for the jammer selection problem, we define two different selection vectors as $\zj_R$ and $\zj_{NR}$, and evaluate $\tilde{f}(\zj_{R}, \bLh_J-\bLd_J^{*})$ and $\tilde{f}(\zj_{NR},\bLh_J-\bLd_J^{*})$.

For the uncertainty region $\mathcal{E}$, each $\lambda_{jk}^{(i)}$ is modeled as  $\lambda_{jk}^{(i)}\in [(1-\epsilon^{(i)})\hat{\lambda}_{jk}^{(i)}, \ (1+\epsilon^{(i)})\hat{\lambda}_{jk}^{(i)}]$ for some $\epsilon^{(i)} \in [0,1]$. Therefore, the eavesdropper selection is based on $(1-\epsilon^{(i)})\hat{\lambda}_{jk}^{(i)}$'s for the robust approach whereas $\hat{\lambda}_{jk}^{(i)}$'s are used for the non-robust approach. It is noted that ${\delta_{jk}^{(i)}}$ in \eqref{eq:robustmodel} can be expressed as ${\delta_{jk}^{(i)} = \epsilon^{(i)}\hat{\lambda}_{jk}^{(i)}}$. If all $\epsilon^{(i)}$'s are not identical (which is commonly the case in practice), we expect performance difference between the robust and non-robust approaches. To that aim, we generate ${N_T = 121}$ realizations of independent uniform random variables distributed in $[0,1]$ for $\epsilon^{(i)}$'s by using MATLAB (the seed is equal to $1$).

For the jammer selection problem, we use a similar setup. For the uncertainty region $\tilde{\mathcal{E}}$, we generate ${N_T = 121}$ realizations of independent uniform random variables distributed in $[0,1]$, denoted as $\kappa^{(i)}$, by using MATLAB (the seed is equal to 2). In this case, the jammer selection is based on the estimate of $\tilde{\lambda}_j^{(i)}$ multiplied with $(1+\kappa^{(i)})$.

\begin{figure}
	\includegraphics[width=0.9\linewidth]{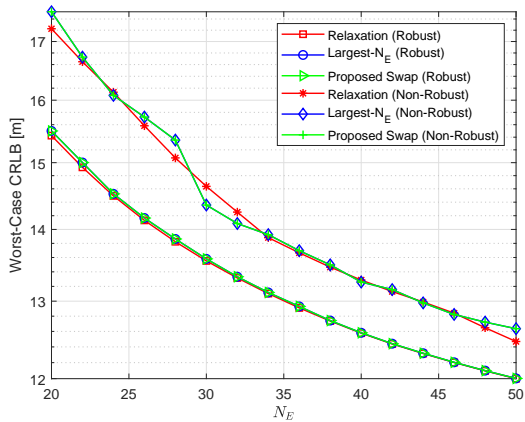}
	\centering
	\caption{{Worst-case CRLB versus $N_E$ when $\sigma^2 = 0.1$ and $N_{\text{swap}}^{\text{max}} = 5$ for the robust eavesdropper selection problem.}}
	\label{fig:RobustEavCRLB}
\end{figure}

\begin{figure}
	\includegraphics[width=0.9\linewidth]{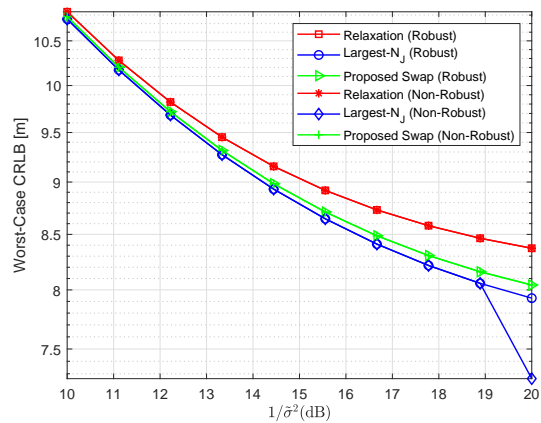}
	\centering
	\caption{{Worst-case CRLB versus $1/\tilde{\sigma}^2$ when $N_J = 5$ and $N_{\text{swap}}^{\text{max}} = 5$ for the robust jammer selection problem.}}
	\label{fig:RobustJamCRLB}
\end{figure}

In Figs.~\ref{fig:RobustEavCRLB} and \ref{fig:RobustJamCRLB}, the worst-case CRLB performances are presented respectively for the eavesdropper selection and the jammer selection problems, considering both the robust and non-robust approaches. In Fig.~\ref{fig:RobustEavCRLB}, as expected, the robust approaches yield lower worst-case CRLBs than the non-robust ones. On the other hand, the robust approach and the non-robust approach perform very similarly in Fig.~\ref{fig:RobustJamCRLB}. {In other words, for this system setup, without having the perfect knowledge of $\tilde{\lambda}_{j}^{(i)}$'s, one can achieve similar CRLB values to those achieved by the robust approach.}


\section{Concluding Remarks}

For wireless source localization networks, the eavesdropper selection, jammer selection, and joint eavesdropper and jammer selection problems have been proposed. Related to the eavesdropper selection problem, a novel CRLB expression has been derived as the performance metric, and its convexity and monotonicity properties have been proved. After relaxing the integer constraints, a convex optimization problem has been obtained and eavesdropper selection algorithms have been proposed. Also, a robust approach has been developed in the presence of uncertainty about system parameters. For the jammer selection problem, a CRLB expression from the literature has been utilized, and its concavity and monotonicity properties have been derived. Similarly, a convex relaxation approach and a robust approach have also been developed for jammer selection. Moreover, the joint eavesdropper and jammer selection problem has been proposed and its relaxed version has been shown to reduce to a convex problem. Various simulation results have illustrated {the benefits of the proposed algorithms in terms of performance and run time. In particular, the performance achieved by the proposed algorithms is very close to the performance bound specified by the relaxed problems, and the corresponding run times are significantly lower than the other alternatives such as the swap algorithm with random initialization and the exhaustive search.}
The results in this manuscript reveal the capabilities of jammer and eavesdropper nodes, which can be useful for designing wireless source localization networks and taking appropriate precautions.

\section*{Appendix}

\subsection{Proof of Proposition 1}

In \cite[Thm.~1]{part1}, the EFIM for estimating the location of a single target node is obtained for synchronized target and anchor nodes. Even though our network model is quite different from the system model described in Section II of \cite{part1}, we benefit from the proof of \cite[Thm.~1]{part1} in the first part of this proof.

In the proof of \cite[Thm.1]{part1}, vector $\bq_{k}$ is defined as $\bq_k=[\cos\phi_k~\sin \phi_k]^\intercal$. We follow the same steps as in that proof 
by replacing vector $\bq_k$ with vector $\bq_{ik}$, which is defined as $\bq_{ik}=[\cos\phi_{ik} ~ \sin\phi_{ik} ~ 1]^\intercal$.\footnote{The reason for using $\bq_{ik}$ instead of $\bq_{k}$ stems from the fact that in our system model, the number of the possible target locations is more than one. Also, the additional term $1$ in $\bq_{ik}$ compared to $\bq_k$ is due to the time offset between the target node and the eavesdropper nodes; i.e., due to the $\Delta_i$ term.} Then, we can obtain the EFIM for $[\xx_i ~ \Delta_i]^\intercal$, denoted by $\mathbf{J}_{e}^{(i)}(\xx_i, \Delta_i,\bz)$,  as follows:
\begin{equation} \label{eq:EFIM1}
	\mathbf{J}_{e}^{(i)}(\xx_i, \Delta_i,\bz) = \begin{bmatrix} K_i(\bz) & D_i(\bz) & C_i(\bz) \\ D_i(\bz) & E_i(\bz) & S_i(\bz) \\ C_i(\bz) & S_i(\bz) & T_i(\bz)
	\end{bmatrix}
\end{equation}
where
\begin{align}
	&K_i(\bz) \triangleq  \sum_{(j,k)\in\mathcal{N}_L^{(i)}}\hspace{-0.1cm}z_k^{E}\lambda_{jk}^{(i)} \cos^2\phi_{ik}, \\
	& E_i(\bz) \triangleq \sum_{(j,k)\in\mathcal{N}_L^{(i)}} \hspace{-0.1cm} z_k^{E} \lambda_{jk}^{(i)} \sin^2\phi_{ik}, \\
	&C_i(\bz) \triangleq  \sum_{(j,k)\in\mathcal{N}_L^{(i)}} \hspace{-0.1cm} z_k^{E} \lambda_{jk}^{(i)} \cos\phi_{ik},\\
	&S_i(\bz)  \triangleq \sum_{(j,k)\in\mathcal{N}_L^{(i)}} \hspace{-0.1cm} z_k^{E} \lambda_{jk}^{(i)} \sin\phi_{ik},
\end{align}
\begin{align}
	&D_i(\bz)  \triangleq \sum_{(j,k)\in\mathcal{N}_L^{(i)}} \hspace{-0.1cm} z_k^{E} \lambda_{jk}^{(i)} \sin\phi_{ik}\cos\phi_{ik}, \\
	& T_i(\bz) \triangleq \sum_{(j,k)\in\mathcal{N}_L^{(i)}} \hspace{-0.1cm} z_k^{E} \lambda_{jk}^{(i)}.
\end{align}

By applying the Schur complement formula to \eqref{eq:EFIM1}, the following expression is obtained:
\begin{align}\nonumber
	\mathbf{J}_{e}^{(i)}(\xx_i,\bz) &= \begin{bmatrix}
		K_i(\bz) & D_i(\bz) \\ D_i(\bz) & E_i(\bz)
	\end{bmatrix} \\
	&-\frac{\begin{bmatrix}
			C_i^2(\bz) & C_i(\bz)S_i(\bz) \\ C_i(\bz)S_i(\bz) & S_i^2(\bz) \end{bmatrix}}{T_i(\bz)}\label{eq:EFIMeq}
\end{align}
Let $\mathbf{J}_1^{(i)}(\xx_i,\bz)$ and $\mathbf{J}_2^{(i)}(\xx_i,\bz)$ be defined as the first and second terms in \eqref{eq:EFIMeq}, i.e.,
\begin{align}\label{eq:EFIMeqT1}
	&\mathbf{J}_1^{(i)}(\xx_i,\bz) \triangleq \begin{bmatrix}
		K_i(\bz) & D_i(\bz) \\ D_i(\bz) & E_i(\bz)
	\end{bmatrix} \\\label{eq:EFIMeqT2}
	&\mathbf{J}_2^{(i)}(\xx_i,\bz)  \triangleq \frac{\begin{bmatrix}
			C_i^2(\bz) & C_i(\bz)S_i(\bz) \\ C_i(\bz)S_i(\bz) & S_i^2(\bz) \end{bmatrix}}{T_i(\bz)}
\end{align}
After some algebra, we derive the following expression from \eqref{eq:EFIMeq}:
\begin{align}
	&\tr\big\{\big(\mathbf{J}_{e}^{(i)}(\xx_i,\bz)\big)^{-1} \big\}  = \nonumber \\
	&\frac{2\sum_{(u,k)\in\mathcal{N}_L^{(i)}} \sum_{(v,l)\in\mathcal{N}_L^{(i)}} p_{k,l}^{(i)} z_k^E z_l^E \lambda_{uk}^{(i)} \lambda_{vl}^{(i)} }{\underset{(u,k)\in\mathcal{N}_L^{(i)}}\sum \underset{(v,l)\in\mathcal{N}_L^{(i)}}\sum \underset{(s,m)\in\mathcal{N}_L^{(i)}}\sum q_{k,l,m}^{(i)} z_k^E z_l^E z_m^E \lambda_{uk}^{(i)} \lambda_{vl}^{(i)} \lambda_{sm}^{(i)}} \label{eq:closedform2}
\end{align}
where
\begin{align}\nonumber
	q_{k,l,m}^{(i)} &=\cos\phi_{ik} \sin\phi_{il}  \sin(\phi_{il}-\phi_{ik})
	\\\nonumber
	&- \cos\phi_{ik}\sin\phi_{im}\sin(\phi_{il}-\phi_{ik}) \\
	&- \cos\phi_{ik} \cos\phi_{il} \sin\phi_{im}  (\sin\phi_{im}-\sin\phi_{ik}).
\end{align}
Based on the trigonometric identity,
\begin{equation*}
	\sin a + \sin b - \sin (a+b) = 4 \sin\bigg(\frac{a}{2}\bigg)\sin\bigg(\frac{b}{2}\bigg) \sin\bigg(\frac{a+b}{2}\bigg)
\end{equation*}
we obtain the following relation:
\begin{align}\nonumber
	&q_{k,l,m}^{(i)} + q_{k,m,l}^{(i)}  + q_{l,k,m}^{(i)}  + q_{l,m,k}^{(i)}  + q_{m,l,k}^{(i)}  + q_{m,k,l}^{(i)}
	\\\label{eq:trigid}
	&=
	16 p_{k,l}^{(i)}  p_{l,m}^{(i)}  p_{m,k}^{(i)}\,.
\end{align}
Then, we can rearrange the denominator of  \eqref{eq:closedform2} as follows:
\begin{align}
	&\sum_{(u,k)\in\mathcal{N}_L^{(i)}} \sum_{(v,l)\in\mathcal{N}_L^{(i)}} \sum_{(s,m)\in\mathcal{N}_L^{(i)}}  q_{k,l,m}^{(i)} z_k^E z_l^E z_m^E \lambda_{uk}^{(i)} \lambda_{vl}^{(i)} \lambda_{sm}^{(i)}  \stackrel{(a)} = \nonumber \\
	&16\hspace{-0.3cm}\sum_{(u,k)\in\mathcal{N}_L^{(i)}} \underset{l>k}{\sum_{(v,l)\in\mathcal{N}_L^{(i)}}}
	\underset{m>l}{\sum_{(s,m)\in\mathcal{N}_L^{(i)}}} \hspace{-0.1cm} p_{k,l}^{(i)}  p_{l,m}^{(i)}  p_{m,k}^{(i)}  z_k^E z_l^E z_m^E \lambda_{uk}^{(i)} \lambda_{vl}^{(i)} \lambda_{sm}^{(i)}  \nonumber\\
	& \stackrel{(b)}= \frac{8}{3} \hspace{-0.4cm} \sum_{(u,k)\in\mathcal{N}_L^{(i)}} \hspace{-0.05cm}\sum_{(v,l)\in\mathcal{N}_L^{(i)}} \hspace{-0.05cm} \sum_{(s,m)\in\mathcal{N}_L^{(i)}} \hspace{-0.4cm}p_{k,l}^{(i)}  p_{l,m}^{(i)}  p_{m,k}^{(i)}  z_k^E z_l^E z_m^E \lambda_{uk}^{(i)} \lambda_{vl}^{(i)} \lambda_{sm}^{(i)}\label{eq:finalExp}
\end{align}
where $(a)$ follows from \eqref{eq:trigid}, and $(b)$ is due to the symmetry in the summand term, $p_{k,l}^{(i)}  p_{l,m}^{(i)}  p_{m,k}^{(i)}  z_k^E z_l^E z_m^E \lambda_{uk}^{(i)} \lambda_{vl}^{(i)} \lambda_{sm}^{(i)}$.
By replacing the denominator of \eqref{eq:closedform2} with the final expression in \eqref{eq:finalExp}, the CRLB expression in \eqref{eq:closedform}--\eqref{eq:rtil} is obtained.\hfill $\blacksquare$

\subsection{Proof of Lemma 1}

As $w_i \geq 0$ in \eqref{eq:fz}, the aim is to show that $\tr\big\{\big(\mathbf{J}_{e}^{(i)}(\xx_i,\bz)\big)^{-1} \big\}$ is non-increasing in $\bz$. Since $\tr\big\{\big(\mathbf{J}_{e}^{(i)}(\xx_i,\bz)\big)^{-1} \big\}$ is non-increasing with respect to $\mathbf{J}_{e}^{(i)}(\xx_i,\bz)$, it is sufficient to prove the following implication:
\begin{equation}
	\bz \succeq \bw \implies \mathbf{J}_{e}^{(i)}(\xx_i,\bz)\succeq \mathbf{J}_{e}^{(i)}(\xx_i,\bw)
\end{equation}
In other words, from \eqref{eq:EFIMeq}, we must prove that
\begin{equation*}
	\mathbf{J}_1^{(i)}(\xx_i,\bz)-\mathbf{J}_1^{(i)}(\xx_i,\bw)-\mathbf{J}_2^{(i)}(\xx_i,\bz)+ \mathbf{J}_2^{(i)}(\xx_i,\bw)  \succeq 0
\end{equation*}
It is noted that for any $\by = [y_1 ~ y_2]^\intercal\in\mathbb{R}^{2}$, the following equalities hold:
\begin{align} \label{eq:quadform1}
	&\by^\intercal \mathbf{J}_1^{(i)}(\xx_i,\bz) \by = y_1^2K_i(\bz) + 2y_1y_2D_i(\bz) + y_2^2E_i(\bz) , \\
	&\by^\intercal \mathbf{J}_2^{(i)}(\xx_i,\bz) \by = \frac{\big(y_1C_i(\bz)+y_2S_i(\bz)\big)^2}{T_i(\bz)}\,\cdot\label{eq:quadform2}
\end{align}
Therefore, by combining \eqref{eq:quadform1} and \eqref{eq:quadform2}, the following relation can be obtained:
\begin{equation}
	\by^\intercal (\mathbf{J}_{e}^{(i)}(\xx_i,\bz)- \mathbf{J}_{e}^{(i)}(\xx_i,\bw)) \by = h_i(\bz)-h_i(\bw)
\end{equation}
where
\begin{align}
	h_i(\bz) &\triangleq y_1^2K_i(\bz) + 2y_1y_2D_i(\bz) + y_2^2E_i(\bz) \\ &- \frac{\big(y_1C_i(\bz)+y_2S_i(\bz)\big)^2}{T_i(\bz)}\,\cdot
\end{align}
Hence, it is sufficient to show that $h_i(\bz)$ is a non-decreasing function of $\bz$.
It is noted that
\begin{align}\nonumber
	&\frac{\partial h_i(\bz) }{\partial z_k^E} = \bar{\lambda}_k^{(i)}(y_1\cos\phi_{ik} + y_2\sin\phi_{ik})^2  \\\nonumber
	&- \bar{\lambda}_k^{(i)}\frac{2(y_1C_i(\bz) + y_2S_i(\bz)) (y_1\cos\phi_{ik} + y_2\sin\phi_{ik})}{T_i(\bz)} \\
	& + \bar{\lambda}_k^{(i)}\frac{(y_1C_i(\bz) + y_2S_i(\bz))^2}{T_i(\bz)^2}
\end{align}
where $\bar{\lambda}_k^{(i)}$ is given by $\bar{\lambda}_k^{(i)} = \sum_{j: (j,k)\in\mathcal{N}_L^{(i)}} \lambda_{jk}^{(i)}\geq0$.
Then, via the arithmetic mean-geometric mean inequality, it is seen that $\frac{\partial h_i(\bz) }{\partial z_k^E}\geq 0$ for any $k = 1, 2, \ldots, N$. Therefore, we have the desired conclusion that $f(\bz)$ is non-increasing in $\bz$.\hfill$\blacksquare$

\subsection{Proof of Proposition 2}

As $w_i\geq 0$ for $i = 1, 2, \ldots, N_{T}$ in \eqref{eq:fz}, it is sufficient to prove that $\tr\big\{\big(\mathbf{J}_{e}^{(i)}(\xx_i,\bz)\big)^{-1} \big\}$ is a convex function of $\bz$. It is known that $\tr\{\bX^{-1}\}$ is a convex function of $\bX$ for any positive semi-definite $\bX$ \cite{boyd}. Also, $\tr\{\bX^{-1}\}$  is non-increasing in $\bX$. Therefore, it is sufficient to prove that $\mathbf{J}_{e}^{(i)}(\xx_i,\bz)$ is a concave function of $\bz$.

To explain why this is sufficient, we define function $g$ as $g(\bX) \triangleq \tr\{\bX^{-1}\}$. Then, we are interested in the convexity of $g(\mathbf{J}_{e}^{(i)}(\xx_i,\bz))$ with respect to $\bz$. In other words, we should prove that for any $\nu\in[0,1]$, and $\bz,\bw \in\mathbb{R}^{N}$,
\begin{align}\nonumber
	g(\mathbf{J}_{e}^{(i)}(\xx_i,\nu\bz + (1-\nu)\bw)) \leq &\,\,
	\nu g(\mathbf{J}_{e}^{(i)}(\xx_i,\bz)) \\\label{eq:sufficiency}
	&+ (1-\nu)g(\mathbf{J}_{e}^{(i)}(\xx_i,\bw))
\end{align}
If $\mathbf{J}_{e}^{(i)}(\xx_i,\bz)$ is a concave function of $\bz$, then $\mathbf{J}_{e}^{(i)}(\xx_i,\nu\bz + (1-\nu)\bw)\geq \nu \mathbf{J}_{e}^{(i)}(\xx_i,\bz)+(1-\nu)\mathbf{J}_{e}^{(i)}(\xx_i,\bw)$ holds. Since $g(\cdot)$ is non-increasing and convex in its argument, it then leads to \eqref{eq:sufficiency}.

In order to prove that $\mathbf{J}_{e}^{(i)}(\xx_i,\bz)$ is a concave function of $\bz$,  we should show that for any $\gamma\in[0,1]$ and $\bz,\bw \in\mathbb{R}^{N}$, the following relation is true:
\begin{equation}\label{eq:objective}
	\mathbf{J}_{e}^{(i)}(\xx_i,\gamma\bz + (1-\gamma)\bw) \succeq
	\gamma \mathbf{J}_{e}^{(i)}(\xx_i,\bz)+ (1-\gamma)\mathbf{J}_{e}^{(i)}(\xx_i,\bw).
\end{equation}
Based on the relations in \eqref{eq:EFIMeq}--\eqref{eq:EFIMeqT2}, the inequality in \eqref{eq:objective} can be reduced to the following:
\begin{equation}\label{eq:objective2}
	\gamma \mathbf{J}_{2}^{(i)}(\xx_i,\bz)+ (1-\gamma)\mathbf{J}_{2}^{(i)}(\xx_i,\bw) \succeq
	\mathbf{J}_{2}^{(i)}(\xx_i,\gamma\bz + (1-\gamma)\bw) 	
\end{equation}
since $\mathbf{J}_1^{(i)}(\xx_i,\bz)$ is linear in $\bz$.

It is deduced from \eqref{eq:quadform2} that for proving \eqref{eq:objective2}, it is sufficient to show that
\begin{align}\label{eq:objective3}
	&\gamma \frac{\big(y_1C_i(\bz)+y_2S_i(\bz)\big)^2}{T_i(\bz)} + (1-\gamma) \frac{\big(y_1C_i(\bw)+y_2S_i(\bw)\big)^2}{T_i(\bw)} \nonumber \\
	&\geq  \frac{\big(y_1C_i(\bs)+y_2S_i(\bs)\big)^2}{T_i(\bs)}
\end{align}
where $\bs = \gamma \bz + (1-\gamma) \bw$. By applying the Cauchy-Schwarz inequality to the left-hand-side of \eqref{eq:objective3}, the following inequality is obtained:	
\begin{align}
	&\hspace{-0.2cm}\gamma \frac{\big(y_1C_i(\bz)+y_2S_i(\bz)\big)^2}{T_i(\bz)} + (1-\gamma) \frac{\big(y_1C_i(\bw)+y_2S_i(\bw)\big)^2}{T_i(\bw)} \geq \nonumber \\
	&\hspace{-0.1cm} \frac{\Big(\gamma\big(y_1C_i(\bz)+y_2S_i(\bz)\big) + (1-\gamma)\big(y_1C_i(\bw)+y_2S_i(\bw)\big)\Big)^2}{\Big(\gamma T_i(\bz) + (1-\gamma)T_i(\bw)\Big)} \label{eq:objective4}
\end{align}
As $C_i(\cdot)$, $S_i(\cdot)$, and $T_i(\cdot)$ are linear in their arguments, \eqref{eq:objective4} is actually the same as \eqref{eq:objective3}, which was to be proved. Hence, the desired conclusion in reached.\hfill$\blacksquare$

\subsection{Proof of Lemma 2}

It is sufficient to show that $\tr\big\{\big(\mathbf{J}_{e}^{(i)}(\xx_i,\bz)\big)^{-1} \big\}$ is non-increasing in $\bl_E^{(i)}$ for any $i = 1, 2, \ldots, N_T$ (see \eqref{eq:fz}). As a consequence of Proposition 1, we can immediately observe that $\tr\big\{\big(\mathbf{J}_{e}^{(i)}(\xx_i,\bz)\big)^{-1} \big\}$ is non-increasing in $\bl_E^{(i)}$ if and only if $\tr\big\{\big(\mathbf{J}_{e}^{(i)}(\xx_i,\bz)\big)^{-1} \big\}$ is non-increasing in $\bz$ due to the symmetric expression in \eqref{eq:closedform}. (That is, the elements of $\bl_E^{(i)}$ and $\bz$ affect the expression in \eqref{eq:closedform} in the same manner.) Therefore, via Lemma 1, we obtain the desired result.\hfill $\blacksquare$

\subsection{Proof of Lemma 3}

It is observed from the expression in \eqref{eq:EFIMJammer} that if $\zj \succeq \tww$, then $\tJ_e^{(i)}(\bx_i,\tww) \succeq \tJ_e^{(i)}(\bx_i,\zj)$ holds for any $i = 1, 2, \ldots, {N_T}$. Since the function $\tr\{(\cdot)^{-1}\}$ is non-increasing in its argument and $w_i\geq 0$ for any $i$, it is concluded that $\tilde{f}(\zj)$ in \eqref{eq:tildefz} is non-decreasing in $\zj$.\hfill$\blacksquare$

\subsection{Proof of Lemma 4}

From \eqref{eq:gi}, the second-order derivatives are calculated as
\begin{equation}\label{eq:SecDers}
	\frac{\partial^2 g_{ij}(\zj)}{\partial z_k^{J} \partial z_l^{J}} = \frac{2\tilde{\lambda}_{j}^{(i)} P_k^{J} P_l^{J} \abs{\gamma_{kj}}^2\abs{\gamma_{lj}}^2 }{(\tilde{\sigma}_j^2+\sum_{l=1}^{N}  z_l^{J} P_l^{J} \abs{\gamma_{lj}}^2)^3}\,\cdot
\end{equation}
Define a vector as $\bv_j \triangleq [P_1^{J}\abs{\gamma_{1j}}^2 ~ \ldots ~ P_N^{J}\abs{\gamma_{Nj}}^2]^{\intercal}$ for $j = 1, 2, \ldots, N_A$. Then, for any $\by\in\mathbb{R}^N$, it follows from \eqref{eq:SecDers} that
\begin{equation}
	\by^{\intercal}\nabla^2 g_{ij}(\zj)\by = \frac{2\tilde{\lambda}_{j}^{(i)}}{(\tilde{\sigma}_j^2+\sum_{l=1}^{N} z_l^{J} P_l^{J} \abs{\gamma_{lj}}^2)^3}  \by^{\intercal} \bv_{j} \bv_{j}^{\intercal} \by \geq 0.
\end{equation}
Therefore, $\nabla^2 g_{ij}(\zj)$ is a positive semi-definite matrix; hence, $g_{ij}(\zj)$ is a convex function of $\zj$.\hfill$\blacksquare$

\subsection{Proof of Proposition 3}

As $w_i\geq 0$ for $i = 1,2, \ldots, N_T$, it is sufficient to prove that $\tr\big\{\big(\tJ_e^{(i)}(\bx_i,\zj)\big)^{-1} \big\}$ is a concave function of $\zj$ for any $i$. We know that $\tr\big\{\big(\tJ_e^{(i)}(\bx_i,\zj)\big)^{-1} \big\}$ is concave with respect to $\zj$ if and only if $\tr\big\{-\big(\tJ_e^{(i)}(\bx_i,\zj)\big)^{-1} \big\}$ convex with respect to $\zj$. Hence, two auxiliary functions are defined as follows:
\begin{align}\label{eq:ctildeDef}
	&\tilde{c}:\mathbb{R}^{2\times 2} \to \mathbb{R} \text{ such that } \tilde{c}(\bX) = \tr\{\bX^{-1}\} \\\label{eq:ciDef}
	&c_i: \mathbb{R}^{N}\to \mathbb{R}^{2\times 2} \text{ such that } c_i(\zj) = -\tJ_e^{(i)}(\bx_i,\zj).
\end{align}
Based on the preceding definitions, $\tr\big\{-\big(\tJ_e^{(i)}(\bx_i,\zj)\big)^{-1} \big\} = \tilde{c}(c_i(\zj))$. It is known that $\tilde{c}(\cdot)$ is convex and non-increasing in its argument \cite{boyd}. Thus, it is sufficient to prove that $c_i(\zj)$ is concave with respect to $\zj$, or equivalently, $\tJ_e^{(i)}(\bx_i,\zj)$ is convex with respect to $\zj$.

To that aim, we should prove that for any $\zj, \tww \in\mathbb{R}^{N}$ and $\tilde{\gamma}\in[0,1]$, the following relation holds:
\begin{equation}
	\tilde{\gamma}\tJ_e^{(i)}(\bx_i,\zj)+ (1-\tilde{\gamma})\tJ_e^{(i)}(\bx_i,\tww) \succeq \tJ_e^{(i)}(\bx_i,\tilde{\gamma}\zj+ (1-\tilde{\gamma})\tww).
\end{equation}
For any $\by = [y_1~y_2]^{\intercal}$, it follows from \eqref{eq:EFIMJammer} and \eqref{eq:gi} that
\begin{equation} \label{eq:EFIMJammereq}
	\by^{\intercal}\tJ_e^{(i)}(\bx_i,\zj) \by = \sum_{j\in\mathcal{A}_i^{L}} g_{ij}(\zj) (y_1 \cos \varphi_{ij} + y_2 \sin \varphi_{ij} )^2.
\end{equation}
By combining Lemma 4 and \eqref{eq:EFIMJammereq}, the desired conclusion is reached.\hfill $\blacksquare$

\subsection{Proof of Lemma 5}

It suffices to show that  $\tr\big\{\big(\tJ_e^{(i)}(\bx_i,\zj)\big)^{-1} \big\}$ is non-increasing in $\bl_J^{(i)}$ for any $i = 1,2, \ldots, N_T$, which is evident from \eqref{eq:EFIMJammer}.\hfill$\blacksquare$

\bibliographystyle{IEEEtran}
\bibliography{papercite2}

\end{document}